\documentclass[12pt,preprint]{aastex}

\usepackage{epsfig}
\usepackage{natbib}
\usepackage{wasysym}
\usepackage{subfigure}
\usepackage{verbatim}

\newcommand{\lapprox} {\, \lower3pt\hbox{$\sim$}\llap{\raise2pt\hbox{$<$}}\,}
\newcommand{\gapprox} {\, \lower3pt\hbox{$\sim$}\llap{\raise2pt\hbox{$>$}}\,}

\begin{document}


\title{GLOBAL ENERGETICS OF THIRTY-EIGHT LARGE SOLAR ERUPTIVE EVENTS}

\author{\sc
A.~G.~Emslie\altaffilmark{1},
B.~R.~Dennis\altaffilmark{2},
A.~Y.~Shih\altaffilmark{2},
P.~C.~Chamberlin\altaffilmark{2},
R.~A.~Mewaldt\altaffilmark{3},
C.~S.~Moore\altaffilmark{4},
G.~H.~Share\altaffilmark{5},
A.~Vourlidas\altaffilmark{6}, and
B.~T.~Welsch\altaffilmark{7}
}

\altaffiltext{1} {Department of Physics \& Astronomy, Western Kentucky University, Bowling Green, KY 42101 \\ {\tt emslieg@wku.edu}}
\altaffiltext{2} {Code 671, NASA Goddard Space Flight Center, Greenbelt, MD 20771 \\{\tt brian.r.dennis@nasa.gov, albert.y.shih@nasa.gov, phillip.c.chamberlin@nasa.gov}}
\altaffiltext{3} {California Institute of Technology, Pasadena, CA {\tt rmewaldt@srl.caltech.edu}}
\altaffiltext{4} {Center for Astrophysics and Space Astronomy, University of Colorado, Boulder, CO \\{\tt christopher.moore-1@colorado.edu}}
\altaffiltext{5} {Department of Astronomy, University of Maryland, College Park, MD {\tt share@astro.umd.edu}}
\altaffiltext{6} {Code 7663, Naval Research Laboratory, Washington, DC {\tt vourlidas@nrl.navy.mil}}
\altaffiltext{7} {Space Sciences Laboratory, University of California, Berkeley, CA {\tt welsch@ssl.berkeley.edu}}

\begin{abstract}

We have evaluated the energetics of 38 solar eruptive events observed by a variety of spacecraft instruments between February~2002 and December~2006, as accurately as the observations allow.  The measured energetic components include: (1) the radiated energy in the GOES 1 -- 8 \AA\ band; (2) the total energy radiated from the soft X-ray (SXR) emitting plasma; (3) the peak energy in the SXR-emitting plasma; (4) the bolometric radiated energy over the full duration of the event; (5) the energy in flare-accelerated electrons above~20~keV and in flare-accelerated ions above~1~MeV; (6) the kinetic and potential energies of the coronal mass ejection (CME); (7) the energy in solar energetic particles (SEPs) observed in interplanetary space; and (8) the amount of free (nonpotential) magnetic energy estimated to be available in the pertinent active region. Major conclusions include: (1) the energy radiated by the SXR-emitting plasma exceeds, by about half an order of magnitude, the peak energy content of the thermal plasma that produces this radiation; (2) the energy content in flare-accelerated electrons and ions is sufficient to supply the bolometric energy radiated across all wavelengths throughout the event; (3) the energy contents of flare-accelerated electrons and ions are comparable; (4) the energy in SEPs is typically a few percent of the CME kinetic energy (measured in the rest frame of the solar wind); and (5) the available magnetic energy is sufficient to power the CME, the flare-accelerated particles, and the hot thermal plasma.

\end{abstract}

\keywords{Sun: activity -- Sun: coronal mass ejections -- Sun: flares -- Sun: particle emission -- Sun: X-rays, gamma rays}

\section{Introduction}

Solar eruptive events (SEEs), which are comprised of flares and associated coronal mass ejections (CMEs), are the most energetic occurrences in the solar system.  Over a period of tens of seconds to minutes, they can convert upwards of $10^{32}$~ergs of energy carried in non-potential, current-carrying magnetic fields into accelerated particles, heated plasma, and ejected solar material.

While the overall energy involved in a large SEE is not in serious doubt, its partition amongst its component parts has so far been estimated only for a few events. In this paper, we provide the first statistical analysis of energy partition throughout the various manifestations of an SEE, for thirty-eight large events.  We provide this information not only to establish ``typical'' ratios of the energy in various components of the event, but also to provide some idea of the range over which such ratios extend, and we especially point out events in which the strength of one component or another appears to lie outside the norm.  We offer this analysis with the goal of providing useful constraints for modelers of the energy release process(es) involved.

This paper grew out of the energetics working group at the meeting on ``Solar Activity during the onset of Solar Cycle 24'' held in Napa, CA, from December~8 - 12, 2008. It is a continuation of the work begun at the Taos ACE/RHESSI/WIND joint workshop in 2003 that led to the works of \citet{2004JGRA..10910104E, 2005JGRA..11011103E}.   These papers provided the first detailed analysis of most of the components of two well-observed SEEs (the GOES X1.5 event on 2002~April~21 and the X4.8 flare event of 2002~July~23), including the energies in thermal plasma, flare-accelerated electrons and ions, associated CME, and solar energetic particles (SEPs). \citet{2004JGRA..10910104E} showed that, for the two events in question, the energy in the magnetic field was sufficient to power the thermal soft X-ray (SXR) emitting plasma, the flare-accelerated electrons and protons, and the kinetic energy in the CME, and they also provided order-of-magnitude estimates of the partition of the energy amongst these components. Subsequently, \citet{2005JGRA..11011103E} also considered the energy in the optical and EUV continua, and they cautioned that, due to the transfer of one energy component to another (e.g., flare-accelerated electrons $\rightarrow$ thermal plasma $\rightarrow$ SXR emission), care must be taken in summing energetic components to arrive at the total energy released in an SEE.  The present paper is also motivated by the work of \citet{2008AIPC.1039..111M}, which was the first to address the ratio of two energetic components (the CME energy in the rest frame of the solar wind and the energy in SEPs) for a statistically significant number of well-observed events.

The basic objective of the paper is to conduct a statistical study of the energy partition into different components for many of the larger SEEs observed during the previous maximum of solar activity, particularly during the period February~2002 through December~2006, the first five years of observations by the Ramaty High Energy Solar Spectroscopic Imager \citep[{\em RHESSI};][]{2002SoPh..210....3L}. The intent is to apply previously proven techniques to determining the global energetics of many more events than the two studied by \cite{2004JGRA..10910104E, 2005JGRA..11011103E}, and, where possible, to apply new techniques to improve the energy estimates.

Our energy estimates come from a wide variety of observations: CME kinetic and potential energies from the Large Angle and Spectrometric COronagraph \citep[LASCO;][]{1995SoPh..162..357B} instrument on the Solar and Heliospheric Observatory (SoHO); energy in flare-accelerated charged particles inferred from the hard X-rays and gamma-rays observed by {\em RHESSI}; energy contained in the SXR-emitting hot plasma from the Geostationary Operational Environmental Satellites (GOES) and {\em RHESSI}; energy in SEPs from the suite of instruments on the Advanced Composition Explorer (ACE) and from GOES, SoHO, the Solar Anomalous and Magnetospheric Particle EXplorer (SAMPEX), and the Solar TErrestrial RElations Observatory (STEREO); and total radiated energy from the Total Irradiance Monitor \citep[TIM,][]{2005SoPh..230...91K} on the SOlar Radiation and Climate Experiment (SORCE).  For weaker events, or where total irradiance measurements are not available, the Flare Irradiance Spectral Model \citep[FISM;][]{2007SpWea...507005C, 2008SpWea...605001C} was used to provide estimates of the bolometric output of a flare based on other measurements.

In Section~\ref{SEEstudied} we present the events studied and review the techniques used to estimate the different component energies of each event. In Section~\ref{comparisons} we present a series of scatter plots of one energy component against another.  While the uncertainties on the individual energy estimates are typically large (often an order of magnitude or greater), these scatter plots, because of the relatively large number of events they contain, nevertheless allow some general conclusions to be reached (Section~\ref{discussion}) about how the energy is partitioned. The spread in the values for the different energy components also gives an idea of the uncertainties in the measured parameters and the range of flare intensities of the selected events. These plots also allow for the identification of a few ``outlier'' events (Section~\ref{outliers}) that indicate either larger measurement uncertainties or distinctly different energy partitioning for those events. We summarize the results in Section~\ref{conclusions}, which also provides suggestions for future work.

\section{Component Energies of the Solar Eruptive Events}
\label{SEEstudied}

The events studied are listed in Table~\ref{tbl-1}.  They include the largest SEP events observed after February 2002, when {\em RHESSI} was launched, excluding those events beyond the West limb (for which no reliable active region identification can be made) and those events located from E60$^\circ$ to E90$^\circ$ (for which the evaluation of the SEP energy is highly uncertain -- see Section~\ref{SEP}). They also include the two events studied by \citet{2004JGRA..10910104E}, which appear as Events \#2 and 6 in Table~\ref{tbl-1}.  Additional events include all flares for which {\em RHESSI} detected significant ($>$4$\sigma$) emission in the 2.223~MeV neutron-capture gamma-ray line \citep{2009PhDT.......427S, 2009ApJ...698L.152S}. This, plus the inclusion of an intriguing behind-the-limb event with a strong CME on 2002~July~20 (Event~\#5 in Table~\ref{tbl-1}), resulted in a total of 38~events for study. As permitted by the available data, estimates were made of the following energies for each of the 38 events:
\begin{enumerate}
\item Radiated energy in the GOES 1 -- 8 $\mbox{\AA}$ band;
\item Total radiated energy from the SXR-emitting plasma;
\item Total (bolometric) radiated output;
\item Peak thermal energy of the SXR-emitting plasma;
\item Energy in flare-accelerated electrons;
\item Energy in flare-accelerated ions;
\item CME kinetic energy in the rest frame of the Sun;
\item CME kinetic energy in the solar-wind rest frame;
\item CME gravitational potential energy;
\item Energy in SEPs; and
\item Free (nonpotential) magnetic energy in the active region.
\end{enumerate}

It is important to keep in mind the differences among the first four items on this list. They are all related, but are included separately since they can each be estimated independently, and indeed relatively straightforwardly, from the available measurements, and since collectively they provide significant information on the thermal energy of each flare, how it is distributed in temperature, and when it is generated and released.  Further details on these four items are provided in subsections~\ref{thermal} to~\ref{peak}.  Broadly speaking, the first item is the energy radiated in the narrow GOES band from 1 -- 8 $\mbox{\AA}$, obtained directly from background-subtracted data (Section~\ref{thermal}).  The second is the energy radiated over {\it all} wavelengths (including the 1 -- 8 $\mbox{\AA}$ band) from the hot SXR-emitting plasma, and is a quantity inferred from the plasma parameters (emission measure and temperature) revealed by the GOES 1 -- 8 $\mbox{\AA}$ measurements.  The third item is the total energy radiated over all wavelengths from {\it all} components of the flare at {\it all} temperatures (including that from the SXR-emitting plasma); in some cases this is directly observed and in some cases inferred from modeling of the emission in select wavelength ranges -- see Section~\ref{Bolometric}. The fourth item does not specify a radiated energy at all, but rather the peak thermal energy content of the hot SXR-emitting plasma; this quantity is inferred (Section~\ref{peak}) from the parameters of spectroscopic fits to {\em RHESSI} data.  It is important to realize \citep{2005JGRA..11011103E} that these four components are not separate flare energy components and therefore cannot be summed together to obtain a total flare thermal energy.

We have not evaluated energy losses from the SXR-emitting plasma by thermal conduction.  However, we would note that conductive transfer of thermal energy from hot SXR-emitting plasma into the relatively cool chromospheric plasma will generally result in the thermal energy content of the hot SXR-emitting plasma being quite effectively transported to, and ultimately radiated away by, such relatively cool plasma, one contribution to the total (bolometric) radiated output -- the third item on the list. Further, for the events considered here there is little observational data available on the energy contained in turbulence and directed mass motions of thermal plasma, components that may well contain energies comparable to the thermal energy of the SXR-emitting plasma \citep[see, e.g.,][]{1992PASJ...44L..95D}.

The data on all the component energies are summarized in Table~\ref{tbl-1}.
GOES SXR data are available for all the events.  However, because of missing or inadequate data or limited instrument sensitivities, definitive energy estimates for all the energy components listed above are available for only six events (Events \#13, 14, 20, 23, 25, and 38). As mentioned above, Event \#5 was located behind the limb; thus only a ``plausible'' magnetic energy estimate (not included in Table~\ref{tbl-1}) could be obtained from observations of the most likely responsible active region once it had moved onto the solar disk, and the listed radiated energies are lower limits.

\begin{deluxetable} {ccrlccrcrrrrrrr}
    \tablewidth{0pt}
    \tabletypesize{\scriptsize}
    \tablecaption{Event List with Component Energies ($\times$ $10^{30}$ ergs)\label{tbl-1}}
    \tablehead{
    \colhead{No.} & \colhead{Date\tablenotemark{*}} & \colhead{Time\tablenotemark{**}} & \colhead{Class} & \colhead{SXR\tablenotemark{1}}& \colhead{T-rad\tablenotemark{2}} & \colhead{Bol\tablenotemark{3}} & \colhead{Peak\tablenotemark{4}} & \colhead{Elec\tablenotemark{5}} & \colhead{Ion\tablenotemark{6}} & \colhead{KE\tablenotemark{7}} & \colhead{SW\tablenotemark{8}} & \colhead{PE\tablenotemark{9}} & \colhead{SEP\tablenotemark{10}} & \colhead{Mag\tablenotemark{11}}}

    \startdata

    1     & 02/02/20 & 05:52  & M5.1  & 0.043  & 1.2   & 13    & \nodata  & \nodata   & \nodata  & 17      & 5.6     & 6.3     & 0.13     & 1200    \\
    2     & 02/04/21 & 00:43  & X1.5  & 1.2   & 38    & 150   & 13       & 20        & \nodata  & 230     & 160     & 5.0     & 23       & 660     \\
    3     & 02/05/22 & 03:18  & C5.0  & 0.048  & 5.6   & 9     & \nodata  & \nodata   & \nodata  & 84      & 45      & 10      & 2.7      & 260     \\
    4     & 02/07/15 & 19:59  & X3.0  & 0.31  & 6.4   & 44    & $>$2.2      & $>$3.6       & \nodata  & 160     & 76      & 10      & 3.8      & 1500    \\
    5     & 02/07/20 & 21:04  & X3.3\tablenotemark{\dag}& $>$1.5& $>$26 & $>$210   & \nodata  & \nodata   & \nodata  & 260     & 170     & \nodata & \nodata  & \nodata \\
    6     & 02/07/23 & 00:18  & X4.8  & 1.2   & 19    & 150   & 2.5      & 32        & 39       & 260     & 150     & 20      & $<$30    & 2000 \\
    7     & 02/08/24 & 00:49  & X3.1  & 1.1   & 24    & 160   & 5.9      & 11        & \nodata  & 210     & 130     & 16      & 3.9      & 2500    \\
    8     & 02/11/09 & 13:08  & M4.6  & 0.11   & 5.0   & 8     & 1.3      & 60        & \nodata  & 180     & 110     & 20      & 0.51     & 550     \\
    9     & 03/05/27 & 22:56  & X1.4  & 0.16  & 3.6   & 16    & 2.8      & 7.4       & 0.19      & \nodata & \nodata & \nodata & \nodata  & 260 \\
    10    & 03/06/17 & 22:27  & M6.9  & 0.21  & 4.6   & 17    & 2.4      & 4.6       & 6.7      & \nodata & \nodata & \nodata & \nodata  & 140 \\
    11    & 03/10/26 & 17:21  & X1.2  & 1.2   & 31    & 88    & \nodata  & \nodata   & \nodata  & 240     & 130     & 32      & 0.75     & 1700 \\
    12    & 03/10/28 & 09:51  & X17   & 4.4   & 68    & 362\tablenotemark{\ddagger} & $>$19     & $>$56     & $>$190   & 1200    & 850     & 63      & 43       & 2900 \\
    13    & 03/10/29 & 20:37  & X10   & 1.9   & 31    & 137\tablenotemark{\ddagger} & 11     & 110       & 30       & 340     & 220     & 25      & 9.7      & 2900 \\
    14    & 03/11/02 & 17:03  & X8.3  & 1.8   & 24    & 130   & 9.3      & 130       & 68       & 270     & 200     & 10      & 9.3      & 2800 \\
    15    & 03/11/03 & 09:43  & X3.9  & 1.1   & 17    & 97    & 2.4      & 120       & 3.1      & \nodata & \nodata & \nodata & \nodata  & 780 \\
    16    & 03/11/04 & 19:29  & X28   & 4.8   & 72    & 426\tablenotemark{\ddagger} & $>$3.1& $>$21    & \nodata & 610     & 410     & 25       & 5.3      & 2800 \\
    17    & 04/07/15 & 18:15  & X1.6  & 0.16  & 4.1   & 8     & 0.93     & 42        & $<$0.1  & \nodata & \nodata & \nodata & \nodata  & 820 \\
    18    & 04/07/25 & 05:39  & M7.1  & 0.069  & 1.3   & 10    & \nodata  & \nodata   & \nodata  & \nodata & \nodata & \nodata & 2.9      & 2300 \\
    19    & 04/11/07 & 15:42  & X2.0  & 0.32  & 5.0   & 56    & 3.0      & 43        & \nodata  & 220     & 130     & 25      & 4.2      & 610 \\
    20    & 04/11/10 & 01:59  & X2.5  & 0.32  & 7.7   & 15    & 2.0      & 20        & 3.4      & 230     & 180     & 16      & 2.4      & 610 \\
    21    & 05/01/15 & 00:22  & X1.2  & 0.23  & 4.7   & 23    & 5.0      & 32        & \nodata  & \nodata & \nodata & \nodata & \nodata  & 1500 \\
    22    & 05/01/15 & 22:25  & X2.6  & 1.3   & 22    & 78    & 7.1      & 63        & 15       & 730     & 540     & \nodata & \nodata  & 1600 \\
    23    & 05/01/17 & 06:59  & X3.8  & 1.8   & 34    & 150   & 17       & 48        & 13       & 1000    & 730     & 50      & 11       & 1600 \\
    24    & 05/01/19 & 08:03  & X1.3  & 0.43  & 7.0   & 54    & 5.9      & 82        & 29       & \nodata & \nodata & \nodata & \nodata  & 1600 \\
    25    & 05/01/20 & 06:36  & X7.1  & 2.9   & 43    & 150   & 10       & 25        & 120      & 15 - 79   & 7.8 - 61  & 2.0     & 7.8      & 1600 \\
    26    & 05/05/13 & 16:13  & M8.0  & 0.44  & 14    & 49    & 3.1      & 13        & \nodata  & 39      & 22      & 4.0     & 7.3      & 400 \\
    27    & 05/07/14 & 10:16  & X1.2  & 0.64  & 12    & 87    & 4.3      & 24        & \nodata  & 100     & 66      & 6.3     & 2.9      & 310 \\
    28    & 05/07/27 & 04:33  & M3.7  & 0.16  & 4.5   & 30    & 1.3      & 12        & \nodata  & 100     & 62      & 10      & \nodata  & 310 \\
    29    & 05/08/22 & 16:46  & M5.6  & 0.34  & 9.8   & 35    & 3.2      & 6.3       & \nodata  & 110     & 76      & 10      & 6.4      & 390 \\
    30    & 05/08/25 & 04:31  & M6.4  & 0.050  & 1.2   & 11    & 1.1      & 16        & $<$1.9   & \nodata & \nodata & \nodata & \nodata  & 110 \\
    31    & 05/09/07 & 17:17  & X17   & 4.9   & 68    & 322\tablenotemark{\ddagger} & $>$5.6   & $>$10    & $>$0.7  & \nodata & \nodata & \nodata & \nodata & 1400 \\
    32    & 05/09/09 & 19:13  & X6.2  & 3.1   & 44    & 250   & $>$7.9      & $>$120    & $>$1.7   & \nodata & \nodata & \nodata & \nodata  & 1300 \\
    33    & 05/09/10 & 21:30  & X2.1  & 0.99  & 17    & 82    & 6.0      & 13        & 1.0      & \nodata & \nodata & \nodata & \nodata  & 1300 \\
    34    & 05/09/13 & 19:19  & X1.5  & 1.1   & 25    & 85    & \nodata  & \nodata   & \nodata  & 330     & 200     & 32      & 3.0      & 1400 \\
    35    & 05/09/13 & 23:15  & X1.7  & 0.23  & 4.7   & 21    & 2.3      & 32        & $<$0.1   & \nodata & \nodata & \nodata & \nodata  & 1400 \\
    36    & 06/12/05 & 10:18  & X9.0  & 1.4   & 19    & 92    & $>$5.1      & $>$360    & $>$4.5   & \nodata & \nodata & \nodata & \nodata  & 400  \\
    37    & 06/12/06 & 18:29  & X6.5  & 1.1   & 18    & 59\tablenotemark{\ddagger} & 6.8     & 40        & 36       & \nodata & \nodata & \nodata & \nodata & 410 \\
    38    & 06/12/13 & 02:14  & X3.0  & 1.1   & 17    & 75    & 4.8      & 13        & 14       & 74      & 44      & 6.3     & 3.2      & 570 \\
    \enddata
\par
\begin{flushleft}
$^*$In yy/mm/dd format. $^{**}$ GOES start time (UT). \newline
$^1$Radiated energy in the GOES 1 -- 8 $\mbox{\AA}$ band. $^2$Total radiated energy from the SXR-emitting plasma.
\newline $^3$Bolometric radiated energy. $^4$Peak thermal energy of the SXR-emitting plasma.
\newline $^5$Energy in flare-accelerated electrons. $^6$Energy in flare-accelerated ions.
\newline $^7$CME kinetic energy in the rest frame of the Sun. $^8$CME kinetic energy in solar-wind rest frame.
\newline $^9$CME gravitational potential energy. $^{10}$Energy in SEPs. $^{11}$Nonpotential magnetic energy in the active region.
\newline $^{\dag}$Behind-the-limb event. $\ddagger$Bolometric irradiance directly measured with TIM -- see Table \ref{tbl-TIM}.
\end{flushleft}

\end{deluxetable}

\subsection{Radiated Energy from Hot Plasma}
\label{thermal}

For each event, we estimated the time-integrated SXR and total radiated energies from the hot SXR-emitting plasma -- the columns labeled `SXR' and `T-rad', respectively, in Table~\ref{tbl-1}.  Fluxes in W~m$^{-2}$ for the 1--8~$\mbox{\AA}$ and 0.5--4~$\mbox{\AA}$ bands are provided by one of NOAA GOES satellites every 3~s.  The total emission in the 1--8~$\mbox{\AA}$ band (`SXR') is obtained simply by summing the background-subtracted fluxes over the duration of the flare, from the GOES start time (given by NOAA and listed in Table~\ref{tbl-1}) to the time when the flux had decreased to 10\% of the peak value. The background that was subtracted was taken as the lowest flux in the hour or so before and/or after the flare.

To calculate the radiated energy from the hot plasma, we used the measured GOES SXR fluxes in a manner similar to that described in \citet{2004JGRA..10910104E}, specifically using the IDL GOES Workbench available in SolarSoftware (SSW).  This allows us to obtain a consistent set of values for all events since GOES, unlike {\em RHESSI}, has full coverage for all events. This calculation assumes that the hot plasma at any given time is isothermal; the temperature and emission measure are calculated from the two-channel GOES data using the relations given by \citet{2005SoPh..227..231W}. Using the thus-inferred emission measure and temperature, and the optically thin radiation loss rate vs.\ temperature function (for coronal abundances and \citet{1998A&AS..133..403M} ionization equilibria) taken from the CHIANTI database \citep{1997A&AS..125..149D, 2009A&A...498..915D}, we used the IDL procedure \emph{rad\_loss}, available in SSW,  to calculate, for each 3-s time interval, the energy radiated from the SXR-emitting plasma over all wavelengths. Finally, we summed the radiated energies over the duration of the flare (from the GOES start time until the 1--8~$\mbox{\AA}$ flux decreased to 10\% of its peak value) to produce the total radiated energy given in the column labeled `T-rad' in Table~\ref{tbl-1}. Significant energy could be radiated after this nominal end of the flare, particularly if there is a ``second phase'' that, according to \cite{2011ApJ...739...59W} and \cite{2011ApJ...731..106S}, can release an amount of energy that is similar to that released in the initial phase. Generally, however, the values quoted in Table~\ref{tbl-1} should include more than 50\% of the energy radiated by the SXR-emitting plasma.

\begin{deluxetable}{cccccccc}

    \tablewidth{0pt}
   \tabletypesize{\scriptsize}
    \tablecaption{TIM and FISM Bolometric Energies ($10^{30}$ ergs)
    \label{tbl-TIM}}

    \tablehead{
     \colhead{Event}& \colhead{Date} & \multicolumn{4}{c}{\hrulefill\ TIM\ \hrulefill} & \colhead{FISM} & \colhead{Difference} \\
      No. & & \colhead{Total\tablenotemark{1}} & \colhead{Uncertainty\tablenotemark{1}} & \colhead{Revised estimate} & \colhead{Corrected\tablenotemark{2}}& \colhead{Corrected\tablenotemark{2}} & \colhead{$\mathrm{(TIM-FISM)/TIM}$}
    }
     \startdata

     12 & 28-Oct-2003 & 600 & 39\% & 362 & 362 & 310 & 14\% \\
     13 & 29-Oct-2003 & 240 & 86\% & 137 & 137 & 128 & 7\% \\
     16 &  4-Nov-2003 & 260 & 65\% & 142 & 426 & 447 & -5\% \\
     31 &  7-Sep-2005 & 300 & 71\% & 150 & 322 & 266 & 17\% \\
     37 &  6-Dec-2006 & \nodata & \nodata & 46\tablenotemark{3}  & 59  & 82  & -39\%

     \enddata

    \tablenotetext{1}{\cite{2006JGRA..11110S14W}}
    \tablenotetext{2}{Corrected for limb darkening}
    \tablenotetext{3}{Moore et al. (2012, in preparation)}

\end{deluxetable}

\subsection{Bolometric Irradiance}
\label{Bolometric}

Estimates of the bolometric irradiance, the total energy radiated from the flare integrated across the entire solar spectrum, for each of the events are provided in the column labeled `Bol' in Table \ref{tbl-1}. For five of the events listed in Table \ref{tbl-1}, the bolometric irradiance was measured directly by the Total Irradiance Monitor \citep[TIM,][]{2005SoPh..230...91K}
onboard SORCE as an increase in the total solar irradiance above the (highly variable) pre- and post flare background levels.  Total flare irradiance values were reported in this manner for Events~\#12, 13, 16, and 31 by \cite{2006JGRA..11110S14W}, and the bolometric irradiance for event \#37 will be reported by Moore et~al. (2012, in preparation). Both previously published values and revised estimates made for this paper are listed in Table~\ref{tbl-TIM}.  A final correction factor (see Table~\ref{tbl-TIM}) was then applied to allow for limb-darkening absorption when the path to the observer becomes optically thick at some wavelengths; the value of this factor can be up to $\sim 3.0$ -- see Equation~(2) in \cite{2006JGRA..11110S14W}.

To complement these direct measurements and so provide a consistent set of bolometric values for all of the events in Table~\ref{tbl-1}, estimates from the Flare Irradiance Spectral Model \citep[FISM;][]{2007SpWea...507005C, 2008SpWea...605001C} were used, with various assumptions and corrections as described below. FISM is an empirical model that provides estimates of the total amount of solar radiated energy over a broad wavelength range from 1-1900~\AA\ and over a wide range of time scales from seconds to years. It uses measurements in this wavelength range from the Solar EUV Experiment \citep[SEE;][]{2005JGRA..11001312W}
on the Thermosphere Ionosphere Mesosphere Energetics and Dynamics (TIMED) satellite and the SOLar-STellar Irradiance Comparison Experiment \citep[SOLSTICE;][]{1993JGR....9810667R}
on the Upper Atmosphere Research Satellite (UARS).

For the relatively rapid GOES 1 -- 8 $\mbox{\AA}$ SXR flux variations that occur during a solar flare, different empirical factors appropriate to the rise and decay phases of the flare, respectively, are used to relate the SXR flux to the total radiated energy during those phases. Various daily proxies are also used to represent the more gradual variations in solar irradiance due to active region evolution, solar rotation, and the solar cycle.  The daily pre-flare irradiance spectrum is subtracted from each value to get the radiated energy from the flare alone, and this is then integrated over the duration of the GOES flare to get the total radiated energy in erg~cm$^{-2}$ at the detector. Then, assuming uniform radiation over 2$\pi$ steradians, the total radiated energy from the flare in the 1-1900~\AA\ wavelength range can be calculated, with 1-minute cadence.

The 1-1900~\AA\ solar irradiance is converted to total radiated energy over all wavelengths (the bolometric irradiance) by multiplying by an empirical conversion factor of $2.42 \pm 0.31$, determined by comparing the 1-1900~\AA\ solar irradiance with the absolute bolometric intensity for the five flares for which the latter could be measured directly (see Table~\ref{tbl-TIM}).

The uncertainties on the calculated values of the bolometric irradiance listed in Tables~\ref{tbl-1} and~\ref{tbl-TIM} are made up of several parts. The most dominant uncertainty comes from the TIM measurements themselves, and is due to the variations in the total solar irradiance of the non-flaring Sun.  Other contributions to the overall uncertainty are the errors on the FISM estimates of the 1-1900~\AA\ flux, the conversion from UV irradiance to total solar irradiance, and the limb-darkening correction.  The overall uncertainty on the calculated values is $\pm\sim$70\% for those events that are near disk center and $\pm\sim$90\% for the near limb events.\footnote[1]{Because the conversion of the FISM radiated energy to bolometric energy is based on the five events measured directly with TIM, the bolometric energies for these five events derived from the FISM estimate differ (after correcting for limb darkening) by less than $40\%$ from the directly measured values.}

\subsection{Peak Thermal Energy Content of the Hot Plasma}
\label{peak}

The peak thermal energy content of the SXR-emitting plasma (the column labeled `Peak' in Table \ref{tbl-1}) was determined from {\em RHESSI} imaging spectroscopy data.  We first fit the observed {\em RHESSI} hard X-ray spectra with the sum of a single-temperature Maxwellian plus the form expected from a double-power-law electron spectrum (Equation~(\ref{eqn:powerlaw}) in Section~\ref{electrons}). The fit parameters appropriate to both thermal and nonthermal components were determined for each time interval using the forward-fitting method implemented in the OSPEX software package available in SSW. The temperatures and emission measures obtained from {\em RHESSI} in this way tend to agree closely with the corresponding values obtained from the standard GOES data analysis discussed in Section \ref{thermal}. However, somewhat higher temperatures can be obtained because of the {\em RHESSI} coverage to higher energies; indeed, superhot components with reported temperatures as high as $\sim$50~MK may exist in some flares and are not accurately reflected in the GOES thermal analysis \citep{1981ApJ...251L.109L, 2010PhDT.........4C, 2010ApJ...725L.161C}. The thermal function included the line plus continuum components determined using CHIANTI, again with coronal abundances and \citet{1998A&AS..133..403M} ionization equilibria. From these fits, the average temperature $T_0$~(K) and emission measure $EM = \int n_e^2 \, dV$ (cm$^{-3}$) of the thermal plasma were determined every 20~s throughout the flare. (Here $n_e$ is the electron density (cm$^{-3}$) and $V$ is the emitting volume (cm$^3$).)

The thermal energy content $U_{\rm th}$ of the plasma can be calculated from the expression \citep[e.g,][]{1986NASACP...2439..505D}
\begin{equation}
U_{\rm th}  =  3 \, n_e \, kT_0 \, f \, V_{\rm ap}
\simeq\ 4.14 \times 10^{-16} \,\, T_0 \, \sqrt{EM \times f \,
V_{\rm ap}} \,\,~{\rm erg}, \label{u_single}
\end{equation}
where $k$ is Boltzmann's constant, $f$ is the volumetric filling factor and $V_{\rm ap}$ is the apparent volume of the SXR source.  Starting in 2003, SXR images are also available from the GOES Soft X-ray Imagers \citep[SXIs;][]{2004SPIE.5171...65L,2005SoPh..226..283P}; such images could be used to provide estimates of $V_{\rm ap}$, as described in \cite{2004JGRA..10910104E}.  However, both for consistency with the earlier analysis of \citet{2004JGRA..10910104E} (which analyzed events that occurred in 2002, prior to the SXI deployment), and since the parameters $EM$ and $T_0$ in Equation~(\ref{u_single}) are deduced from {\em RHESSI} data, we have chosen to use estimates of $V_{\rm ap}$ that are deduced from {\em RHESSI} images, made using the 3$\sigma$-clean method of \citet{2009ApJ...698.2131D}.  We further take $f$ to be unity, consistent with \cite{2004JGRA..10910104E} and further justified by the recent work of \citet{2012ApJ...755...32G}, who used hard X-ray imaging spectroscopy data of 22 extended-loop events to derive a (logarithmic) mean filling factor $f = 0.20 \times\!\!/\!\div 3.9$ ($1\sigma$ standard error).

 The peak energy values listed in Table~\ref{tbl-1} are the highest values of $U_{\rm th}$ obtained from this analysis, usually at or near the time of the peak GOES flux.

\subsection{Flare-Accelerated Electrons}
\label{electrons}

The energies in flare-accelerated electrons are listed in the column labeled `Elec' in Table~\ref{tbl-1}. They were determined by using the OSPEX algorithm to fit a combined  isothermal-plus-nonthermal function to the measured {\em RHESSI} spatially integrated X-ray spectra.  The nonthermal component was assumed to be bremsstrahlung from energetic electrons with an injected spectrum $F_0(E_0)$ (electrons~cm$^{-2}$~s$^{-1}$~keV$^{-1}$) in the form of a broken power-law:
\begin{equation}
    \label{eqn:powerlaw}
    F_{0}(E_0) = A\left\{
        \begin{array}{ll}
            0 & , E_0 < E_{\rm min} \\
            (E_0/E_{p})^{-\delta_{1}} & ,E_{\rm min} \le E_0 < E_{b} \\
            (E_0/E_{p})^{-\delta_{2}} (E_{b}/E_{p})^{\delta_{2}-\delta_{1}} & ,E_{b} \le E_0 < E_{\rm max}\\
            0 & ,E_{\rm max}\le E_0 \\
        \end{array}
    \right.
    .
\end{equation}
The seven parameters of this model spectrum are the normalization parameter, $A$, the low and high energy cutoffs, $E_{\rm min}$ and $E_{\rm max}$, the break energy $E_{b}$, and the power-law indices $\delta_{1}$ and $\delta_{2}$ below and above the break energy, respectively.  The (arbitrary) value of the pivot energy $E_p$ was fixed at 50~keV. Also, the high energy cutoff $E_{\rm max}$ was fixed at 30~MeV, an energy so high above the energy range of interest ($\lapprox 500$~keV) that it has a negligible effect on the calculated X-ray spectrum and so is equivalent to having no high-energy cutoff at all.

The OSPEX analysis uses a forward-fitting procedure that starts by dividing the flare into multiple time intervals -- here we used 20~s intervals. For each interval, the function \emph{thick2} in SSW is used with a set of starting parameters for the electron spectrum~(\ref{eqn:powerlaw}) used to calculate the X-ray photon spectrum, assuming electron-ion bremsstrahlung in a thick target that is ``cold'' in the sense that the ambient electrons have a mean energy $kT$ that is significantly lower than the lowest energy of the accelerated electrons. In general, consideration must also be given to the ionization state of the target, since the bremsstrahlung efficiency is a factor of $\sim$3 times higher for a fully-ionized plasma than for an un-ionized gas \citep{1973SoPh...28..151B,2003ApJ...595L.123K}. However, since most of the beam energy is in the lower energy electrons that stop higher in the corona, we used parameters appropriate for a fully-ionized plasma to calculate the total nonthermal energy. A more refined calculation is possible using the procedure outlined by \cite{2002SoPh..210..419K}
and \cite{2011ApJ...731..106S}, but no significant difference is expected in the resulting total energy in electrons above $E_{\rm min}$.

The resulting photon spectrum is then folded through the detector response matrix to generate a count-rate spectrum, which is added to the count-rate spectrum calculated for the thermal spectrum discussed in Section \ref{thermal}. Then, through an iterative procedure, we find best-fit values of the parameters describing the electron spectrum~(\ref{eqn:powerlaw}), by minimizing the $\chi^2$ statistic between the calculated and the measured background-subtracted count spectra. The total energy $U_e$ in electrons for a given event is then computed by integrating the best-fit electron energy spectrum above $E_{min}$ for each time interval and summing the results over the duration of the flare, resulting in the values listed in the column labeled `Elec' in Table~\ref{tbl-1}.

In order to obtain the most reliable spectral fits to the {\em RHESSI} data and thus better evaluate the uncertainties in the calculated values of $U_{\rm e}$, we chose to use data from just one of {\em RHESSI}'s nine detectors -- Detector \#4. This particular detector has good energy resolution and sensitivity, which allowed us to apply the most up-to-date corrections for energy resolution and calibration, photospheric albedo, pulse pile-up, and background subtraction that are available with the current analysis software. For the large events studied, the count rates were sufficiently high that selecting just a single detector did not seriously degrade the spectroscopic capability up to the photon energies required to determine the parameters of interest. \cite{2009ApJ...699..968M} have shown that similar best-fit parameter values are determined using different individual detectors (detectors 1, 3, 4, 5, 6, and 9 in their case), which leads to an estimate of the systematic uncertainties in the calculated total energy in electrons of $\sim$20\%. This is negligible compared to the uncertainty arising from the difficulty in establishing the value of the low energy cutoff energy $E_{\rm min}$, an uncertainty that arises because the thermal emission generally dominates the low-energy part of the X-ray photon spectrum up to energies where the effects of a cutoff in the electron spectrum might be detectable. We used the largest value of $E_{\rm min}$ that still gave an acceptable fit (reduced $\chi^2 \simeq 1$). As a result, the values of $U_{\rm e}$ listed in Table \ref{tbl-1} are lower limits to the energy in the nonthermal electrons. Furthermore, because of the steep form of the electron spectra ($\delta_{1} \gapprox4$), these values are particularly sensitive to $E_{\rm min}$, so that the energies in flare-accelerated electrons could be up to an order of magnitude higher than those reported in Table~\ref{tbl-1}.

\subsection{Flare-Accelerated Ions}
\label{ions}

The energies in flare-accelerated ions with energies above 1~MeV are listed in the column labeled `Ion' in Table \ref{tbl-1}. In order to provide a consistent set of values for as many events as possible, the energies were estimated solely from {\em RHESSI} measurements of the fluence (time-integral of the flux, photons~cm$^{-2}$) in the 2.223~MeV neutron-capture gamma-ray line. Our sample of events is primarily based on the studies of \citet{2009PhDT.......427S} and \citet{2009ApJ...698L.152S}, who analyzed {\em RHESSI} flares from 2002 to 2005 that had either 2.223~MeV line emission and/or $>$0.3~MeV electron bremsstrahlung continuum emission.  Of those flares, energies are included only for those that have $>$2$\sigma$ detections of that line, and further only as $4\sigma$ upper limits if below a $4\sigma$ detection. We also include three additional flares that occurred in 2006 (Events \#36, 37 and 38).  We chose a lower energy threshold of 1~MeV because the production of detected nuclear gamma-ray lines from elements such as $^{20}$Ne begins at energies as low as $\sim$3~MeV, and it is therefore evident that the ion spectrum extends down to $\sim$1~MeV, at least in a few large events \citep{1995ApJ...455L.193R, 2000IAUS..195..123R}. The spectral shape is essentially unknown below 1~MeV.

In order to estimate the energy in ions from the 2.223~MeV line fluences, the following steps were taken. The measured fluences of the line were first corrected for attenuation in the solar atmosphere assuming a given depth of production of the photons \citep{1987SoPh..107..351H} and allowing for the flare position on the solar disk. The corrected fluence values were then converted to the proton energy above 30~MeV using conversion factors given by \cite{2007ApJS..168..167M} and \cite{2009PhDT.......427S}. The 30~MeV threshold was used at this stage in the analysis because the 2.223~MeV line is produced by ions with energies $\gapprox 20$~MeV~nucleon$^{-1}$, so that the conversion factors are less dependent on the assumed power-law index of the proton spectrum.

In order to estimate the energy in protons above 1~MeV, an extrapolation is required over one-and-a-half orders of magnitude in proton energy, so that the inferred energy above 1~MeV depends critically on the spectral index used in this extrapolation. For the largest {\em RHESSI} flares, where multiple types of ion-associated gamma-ray emission can be detected and fit simultaneously, the ion power-law spectral indices are found to be typically in the range 3--5, a range of indices consistent with that found in a study of flares observed by the Gamma-Ray Spectrometer on the Solar Maximum Mission \citep{1996AIPC..374..172R}.
Consequently, for the purposes of estimating the total energy in protons, we have assumed a power-law proton spectrum with a single spectral index of 4 that extends down to a lower cutoff energy of 1~MeV. Because of the long ``lever arm'' associated with this extrapolation, an uncertainty in the spectral index of $\pm$1 corresponds to an uncertainty in the total energy content above 1~MeV of about $\pm$1.5~orders of magnitude.

Even under the assumption that the spectra for the various types of ions have the same spectral index and low-energy cutoff, the energy content will also depend on the accelerated particle composition.  The ratio of the energy content in all ions (protons plus $\alpha$-particles and heavier nuclei) to the energy in protons can vary between $\sim$2 and $\sim$6; here we assume that the energy in flare-accelerated ions is three times the energy content in flare-accelerated protons.

For a number of the events, the total energy content of ions listed in Table \ref{tbl-1} is a lower limit because {\em RHESSI} did not see the complete time history as the result of spacecraft night or passage through the South Atlantic Anomaly (SAA).  In addition, there can be other complications that affect the observation or interpretation of the neutron-capture line. The affected events are as follows:

\begin{itemize}

\item Event \#12: {\em RHESSI} missed a significant fraction of the neutron-capture line emission, including the peak, as shown by observations of this flare by \emph{\mbox{INTEGRAL}} \citep{2006A&A...445..725K};

\item Event \#31: {\em RHESSI} observed only $\sim$2~minutes of a significantly longer impulsive phase.  Furthermore, the level of atmospheric attenuation is very uncertain due to this flare's large heliocentric angle, so we use a conservative angle of 80$^{\circ}$ to determine the correction factor;

\item{Events \#32 and 37}: {\em RHESSI} missed the peak of the impulsive emission, and thus possibly a significant fraction of the total emission;

\item{Event \#36}: {\em RHESSI} missed some fraction of the 2.223~MeV emission as it was just coming out of Earth shadow.  {\em RHESSI} observations started at 10:31~UT and the GOES X-ray flare started at $\sim$10:19~UT;

\item{Events \#14, 15, 22, 33, and 38}: {\em RHESSI} likely missed a small fraction of the neutron-capture line emission late in the flare. However, this missing energy is smaller than the other uncertainties in the energy estimates discussed above;

\item{Events \#36, 37, and 38}: By December 2006, {\em RHESSI}'s detectors had reduced gamma-ray sensitivity resulting from accumulated radiation damage, a reduction that is difficult to estimate.

\end{itemize}

\subsection{Coronal Mass Ejection}
\label{CME}

The CME kinetic energies, both in the rest frame of the Sun and in that of the solar wind (for comparison with the SEP energies - see Section \ref{SEPvsCME_KESW}), are listed in the columns labeled `KE' and `SW,' respectively, of Table \ref{tbl-1}. The gravitational potential energies of the CMEs are listed in the column labeled `PE.' These CME energies were estimated from calibrated LASCO images using the procedure detailed in \cite{2010ApJ...722.1522V, 2011ApJ...730...59V}.

Briefly, this procedure consists of the following steps. First, we selected two LASCO images, one containing the CME and the other taken before the event as close in time as possible to the flare with no disturbances or ejecta over the path of the subsequent CME. Next, the images were calibrated (in units of mean solar brightness) and the pre-event image was subtracted from the CME image. The excess brightness revealed by this subtracted image is due to Thompson scattering of photospheric radiation from the excess mass in the CME. This excess brightness can therefore be converted to excess mass of the CME under the usual assumptions that
(1) all of the CME mass is concentrated on the plane of the sky, and (2) the CME material consists of 90\% H and 10\% He \citep{1981SoPh...69..169P, 2000ApJ...534..456V, 2010ApJ...722.1522V}. We used the first assumption because the true three-dimensional distribution of the CME mass along the line of sight is unknown. The second assumption represents an ``average'' coronal composition, since we do not know the height at which the bulk of the CME material originates (other than that it is coronal).

These assumptions together result in a lower limit for the mass.  The uncertainty in the CME mass becomes more significant as the central angle and/or spread of a given CME departs significantly from the plane of the sky. The mass underestimation is about a factor of~2 for CMEs that are $\lapprox 40^{\circ}$ from the sky plane \citep{2010ApJ...722.1522V}.

Other uncertainties in this procedure include exposure time variations between event and pre-event images, improper vignetting correction, solar rotation effects, and the presence of stars in the field of view. Fortunately, such uncertainties can be minimized to a level that is well below that of other factors through proper calibration and careful choice of event and pre-event images.

After obtaining a series of excess masses of the CME as a function of time, we can compute both the total mass of the CME and the position and projected velocity, both for the leading edge and for the center of mass of the CME. From the mass $m$, position $r$, and velocity $V$ we can straightforwardly estimate the total kinetic ($U_{\rm K}= {1 \over 2} m V^2$) and potential ($U_{\Phi} = GM_\odot m[R_\odot^{-1} - r^{-1}]$) energies (here $G$ is the Newtonian gravitational constant and $M_\odot, R_\odot$ are the solar mass and radius, respectively). These values are again lower bounds since both the mass and the speed are projected quantities. \cite{2010ApJ...722.1522V}
estimate that, for CMEs that are far away from the sky plane and that have relatively small widths, the kinetic energy could be as much as a factor of eight times larger than the values derived above; similarly the potential energy could be as much as twice as large for such events. However, for the majority of events, the uncertainties on the quoted energies are within a factor of two. To obtain the kinetic energy in the solar wind rest frame (as an estimate of the energy available for shock acceleration of SEPs; see Section \ref{SEPvsCME_KESW}), we simply subtracted 400 km~s$^{-1}$ from the measured CME speed and recomputed the kinetic energy using the speed in this new reference frame.

\subsection{Solar Energetic Particles (SEPs)}
\label{SEP}

For the majority of the events studied, it is likely that the interplanetary SEPs in the events studied are accelerated by CME-driven shocks. \citep[A possible exception is the 2002~February~20 event, where particles directly accelerated in the flare could dominate; see][]{2010JGRA..11506101C}.
The energy content of the accelerated SEPs, particularly when compared to the kinetic energy of the CME in the solar wind rest frame, is therefore an important measure of the efficiency of SEP production by the CME.

The energy content of SEPs that escape into interplanetary space has been estimated by measuring the energy spectra of electrons from $\sim$0.035 to $\sim$8~MeV, protons from $\sim$0.05 to $\sim$400~MeV~nucleon$^{-1}$, and abundant heavier ions from $\sim$0.05 to $\sim$100~MeV nucleon$^{-1}$. Estimates were made in a number of large events from Solar Cycle 23 using a combination of nine separate instruments.
The proton spectra are based on data from the Ultra-Low Energy Isotope Spectrometer \citep[ULEIS;][]{1998SSRv...86..409M}, and the Electron, Proton, and Alpha Monitor \citep[EPAM;][]{1998SSRv...86..541G} on ACE; from the Proton/Electron Telescope \citep[PET;][]{1993ITGRS..31..557C} on SAMPEX; and from the Energetic Particle Sensors \citep[EPS;][]{1996SPIE.2812..281O} on NOAA's GOES-8 and GOES-11 satellites. Spectra of helium and heavier ions were measured by the Solar Isotope Spectrometer \citep[SIS;][]{1998SSRv...86..357S} on ACE and by ULEIS.  Also used for the 2006 events were two STEREO instruments, the Low Energy Telescope \citep[LET;][]{2008SSRv..136..285M} and High Energy Telescope \citep[HET;][]{2008SSRv..136..391V}. Electron measurements were provided by ACE/EPAM, SAMPEX/PET, and by the Electron Proton Helium INstrument \citep[EPHIN;][]{1995SoPh..162..483M} instrument on SoHO.

For eleven of these events, the energy spectra of H, He, and abundant heavier ions were all fit with common spectral forms that include the double-power-law function of \cite{1993ApJ...413..281B}
and the \citet{1985ApJ...298..400E} spectrum -- a power-law with an exponential cutoff. Examples of energy spectra and both functional forms are given in \cite{2005JGRA..11009S18M, 2012SSRv..tmp...32M} and \cite{2005JGRA..11009S16C}.  For the remainder of the events, the proton energy spectra were fit and the contributions of He and heavier ions were estimated using element abundances measured for these events by ULEIS and SIS. The electron contribution was measured in each of the individual events using either EPAM and PET, or EPAM and EPHIN.

For all of the fluence measurements described above, the instruments were located near-Earth.  As in \cite{2004JGRA..10910104E}, we
used the measured near-Earth fluence spectra, typically integrated over 3~to 5 days, to estimate the energy~cm$^{-2}$ that escaped beyond 1 AU in the form of SEPs.  To obtain this estimate, \cite{2004JGRA..10910104E} corrected for the fact that SEPs can scatter back and forth across 1 AU (providing multiple opportunities to be measured) using correction factors based on simulations by Giacalone (personal communication, 2002).  A similar approach was followed in analyzing the ``Halloween'' events \citep{2005JGRA..11009S18M} and in a subsequent survey of 17 events \citep{2006SSRv..124..303M}.
\cite{2008AIPC.1039..111M}
improved on these estimates in a study of 23~SEP events from 1997-2005 by correcting for the fact that SEPs also lose energy as they scatter on the diverging interplanetary magnetic field (IMF).

For this work, we corrected for both multiple 1-AU crossings and energy loss using new simulations by \cite{2010JGRA..11506101C}
for four species (H, He, O, and Fe) with a range of charge-to-mass ratios. \cite{2010JGRA..11506101C} considered scattering mean free paths $\lambda$ ranging from~0.01~to~1~AU, and also varied the radial and rigidity dependence of $\lambda$. Surprisingly, the source energy required to account for the accelerated particles in these different scattering descriptions varied by less than a factor of $\sim$2. This is apparently because the scattering and energy-loss processes compensate for each other -- the more particles scatter the more often they cross 1 AU, but they also lose more energy in the process.  In this paper we have used their form of $\lambda$ derived from quasi-linear theory \citep[see Equation~(3) in][]{2010JGRA..11506101C}.

To relate the measured near-Earth values of MeV~cm$^{-2}$ to the integrated contribution of SEPs escaping through a 1-AU sphere surrounding the Sun, we need to know how SEPs from a given source location are distributed in longitude and latitude. \cite{2004JGRA..10910104E}
assumed that the SEP fluence at Earth falls off exponentially with e-folding separations of 35$^{\circ}$ for latitude, 45$^{\circ}$ for longitude in western events, and 25$^{\circ}$  for longitude in eastern events.  Since then, \cite{2006ApJ...653.1531L}
have measured the longitudinal distribution of SEPs using two- and three-spacecraft data from the two Helios spacecraft and the Interplanetary Monitoring Platform-8 (IMP-8).  They adopted a Gaussian spatial distribution given by $F = F_{o}~\exp[-(\Phi-\Phi_{0})^{2}/2\sigma^2]$, where $\Phi$ is the longitude of the observer, $\Phi_0$ is located 25$^{\circ}$.8 east of the point of best solar wind connection for a 450 km~s$^{-1}$
solar wind ($\sim$W52$^\circ$), and $\sigma = 38^\circ$.  We use their result for the fluence of 4~- 13 MeV protons and we assume it also applies to latitude differences.  By using this relation with the measured flare location and the near-Earth value for the escaping MeV~cm$^{-2}$, we obtained the source energy required to supply SEPs escaping over a 1-AU sphere centered on the Sun.  The results are tabulated in the `SEP' column of Table~\ref{tbl-1}.  Note that we have limited this study to SEP events with source regions ranging from E60° to W90° in longitude.  Beyond this range the Gaussian longitude distribution adopted by \cite{2006ApJ...653.1531L} drops off very rapidly and the longitude corrections become considerably greater and more uncertain. The typical uncertainty in SEP energy is conservatively estimated to be a factor of three.

\subsection{Nonpotential Energy in the Active Region Magnetic Field}
\label{Magnetic}

It is commonly believed that the fundamental energy source for an SEE lies in current-carrying magnetic fields.  In such a scenario, the free energy available to power the event is the excess ``non-potential'' magnetic energy -- the energy above the minimum-energy, potential (i.e., current-free) field to which the field can relax. The estimated available nonpotential magnetic energies of the active region producing the SEEs are listed in the column labeled `Mag' of Table~\ref{tbl-1}. The estimates were made from full-disk line-of-sight (LOS) magnetograms obtained from the Michelson Doppler Imager \citep[MDI;][]{1995SoPh..162..129S} on SoHO, using the method described by \cite{2009ApJ...705..821W} and outlined below.

The 2002~July~20 event (Event~\#5) was located behind the east limb.  Although its source is therefore uncertain, subsequent active region maps suggest that this event probably occurred in AR~10039, the same source region for the 2002~July~23 event (Event~\#6). Consequently, the estimated non-potential magnetic energy for the latter event is a plausible estimate for the non-potential magnetic energy in the former, and indeed the use of such an estimate is consistent with the procedure used to estimate the non-potential magnetic energy content in other near-limb events (e.g., W72$^\circ$ for 2002~February~20 [Event~\#1] and W84$^\circ$ for 2002~April~21 [Event~\#2]), for which magnetic fields measurements obtained when the pertinent active region was near to disk center were used.  However, because there is still a finite possibility of misidentification of the active region, we have chosen not to use this estimated value either in Table~\ref{tbl-1} or in the pertinent plots of Section~\ref{comparisons}.

Numerous efforts have been undertaken to estimate nonpotential magnetic energies in active regions near disk center. The methods include: (1) using the magnetic virial theorem estimates from chromospheric vector magnetograms \citep{1995ApJ...439..474M, 2005ApJ...623L..53M}; (2)~semi-empirical flux-rope modeling using H$\alpha$ and EUV images with MDI line-of-sight (LOS) magnetograms \citep{2008ApJ...672.1209B}; and (3)~MHD modeling \citep{1995ApJ...439..474M, 1997SoPh..174..311J} and non-potential field extrapolation based upon photospheric vector magnetograms \citep{2008ApJ...679.1629G, 2008ApJ...675.1637S, 2008A&A...484..495T, 2008A&A...488L..71T}.
These methods are labor intensive, and uncertainties in their energy estimates are large. For example, error bars on virial free energy estimates can exceed the potential magnetic energy. Also, there is considerable scatter in estimates from studies that employ several methods to analyze the same data \citep[e.g.,][]{2008ApJ...675.1637S}. A couple of generalizations, however, can be made. Free energies determined by virial methods matched or exceeded the potential field energy, while free energies estimated using other techniques typically amounted to a few tens of percent of the potential field energy.  Published values for free energies in analytic \citep{2006SoPh..235..161S}
and semi-empirical \citep{2008SoPh..247..269M}
fields meant to model solar fields also hover around a few tens of percent of the potential field energy.

We estimated the free (i.e., nonpotential) magnetic energies listed in Table \ref{tbl-1} to be 30\% of the potential magnetic energy determined from MDI full-disk LOS magnetograms. This is believed to be a conservative estimate but it has the advantage that it can be readily determined for most of the events. Some of the events arose from limb active regions, for which simultaneous magnetograms are unavailable. Even if vector magnetograms were available, uncertainties in free energies would still be large.  With published virial free energy estimates ranging to a few times the potential energy, it is possible that the true free energy could exceed our estimates by a factor of $\sim$10.

Apart from two cases where flux was clearly emerging near the time of the event (Events \#29 and \#38), we calculated the potential magnetic energies from magnetograms in which each event's source active region was near the disk's central meridian, assuming a rigid rotation rate of 13$^\circ$~day$^{-1}$.  This means the energy estimates were sometimes made a few days before or after a given event.  Fields were assumed to be radial, so each pixel's line-of-sight field strength $B_{\rm LOS}$ was divided by the cosine of the heliocentric angle between the pixel and the sub-observation point, to generate an estimated radial field, $B_R$.  Using a Mercator projection \citep{2009ApJ...705..821W},
the corrected pixel values were then interpolated onto a 2-D plane.  Next, the scalar potential $\chi$, where $\mathbf{B} = -\nabla \chi$, was determined using a Green's function method.  Finally, the magnetic energy $U_M$ was estimated by integrating $(\chi B_R/8\pi)$ over manually-defined cropping windows that contained each active region. Images of the magnetograms used, as well as deprojected data with cropping windows, are online at {\it http://solarmuri.ssl.berkeley.edu/$\sim$welsch/public/meetings/SADOSC24/}.

In several cases in Table~\ref{tbl-1} (e.g., Events \#22 to 25), the same value of the magnetic energy is given for adjacent events up to five days apart from the same active region. This is because the magnetic energy was estimated from line-of-sight magnetograms taken when the active region was close to disk center. These estimates become increasingly unreliable as the active region moves away from disk center. Thus, although it is very likely that the active region's magnetic fields evolved substantially over the time between events, there is no way to reliably quantify these changes from the available magnetograms. This problem will be alleviated with the now regularly available vector magnetic field measurements from the Helioseismic and Magnetic Imager \citep[HMI;][]{2012SoPh..275..207S} on the Solar Dynamics Observatory (SDO), which can be used to estimate the energy in regions located away from disk center.

\section{Comparisons of Energetic Components}
\label{comparisons}

In this section we present comparisons of the energy contents of the various components discussed above, through a series of figures each showing logarithmic scatter plots (the ``forest'') of the energy content of one component versus the energy content of another, for all events (``trees'') for which data are available for both selected components.  The scatter of the points around the logarithmic centroid is due both to the true range of energies of the selected events and to the often large uncertainties (up to 2.5 orders of magnitude in some cases) in the energy estimates of each component. If the uncertainties are random, then the centroid location gives an indication of the average ratio of the energies of the two components being plotted, and the scatter of the points about the centroid provides a measure of the overall uncertainty in that ratio. Any ``outlier'' point indicates an anomalous event, which could simply identify an unusually large or small event, or which could reveal intrinsic differences in the distribution of energies between the different components or some error in the energy estimates for the event in question.

\begin{figure}[pht]
\begin{center}
    \subfigure[][GOES 1--8 $\mbox{\AA}$ band vs.\ the total energy radiated from the SXR-emitting plasma.]{\includegraphics[width=0.40\textwidth, bb=0 0 400 360]{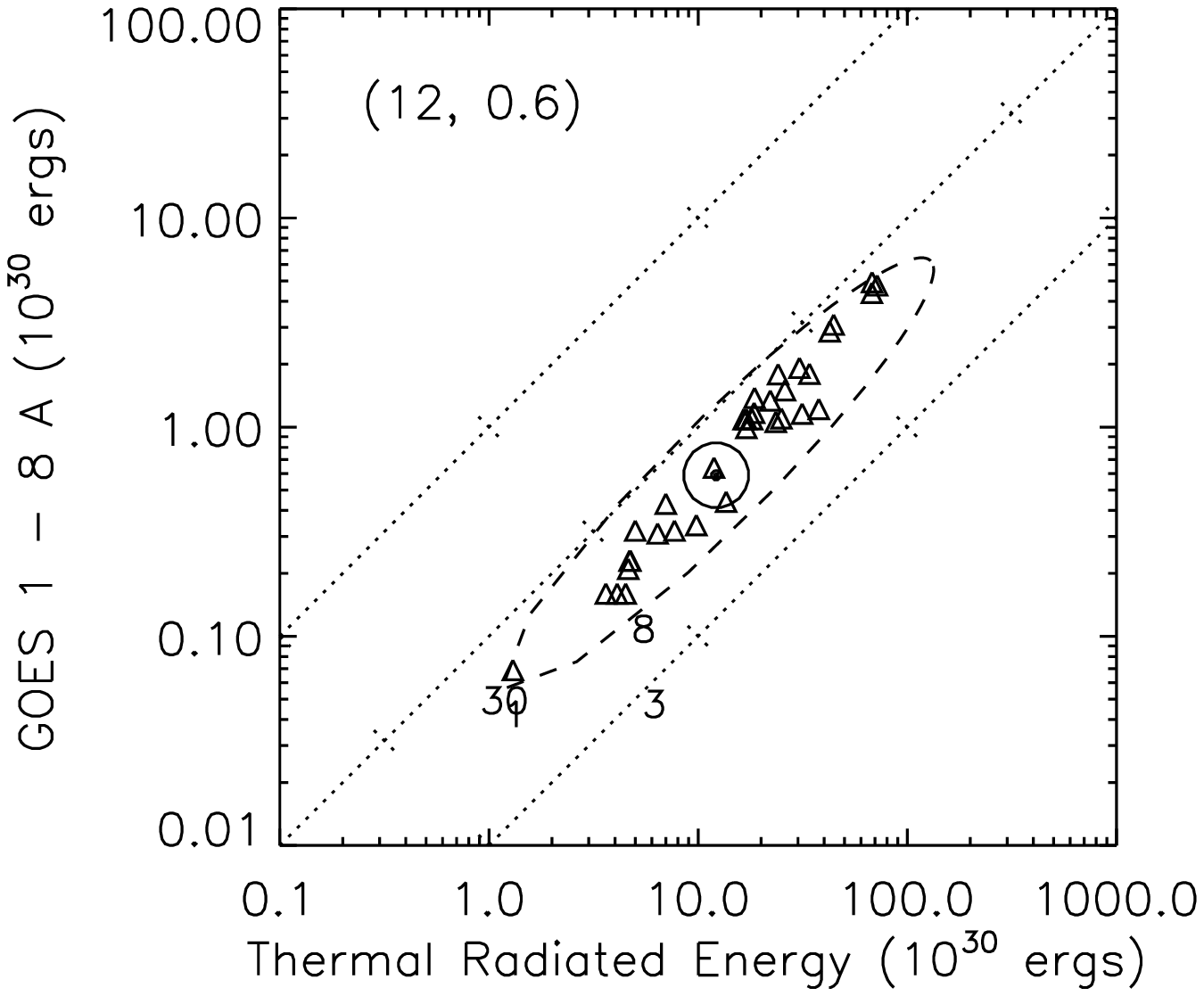}\label{1a}} \quad
    \subfigure[][Total energy radiated from the SXR-emitting plasma vs.\ the peak thermal \mbox{energy} content of that plasma.]{\includegraphics[width=0.40\textwidth, bb=0 0 400 360]{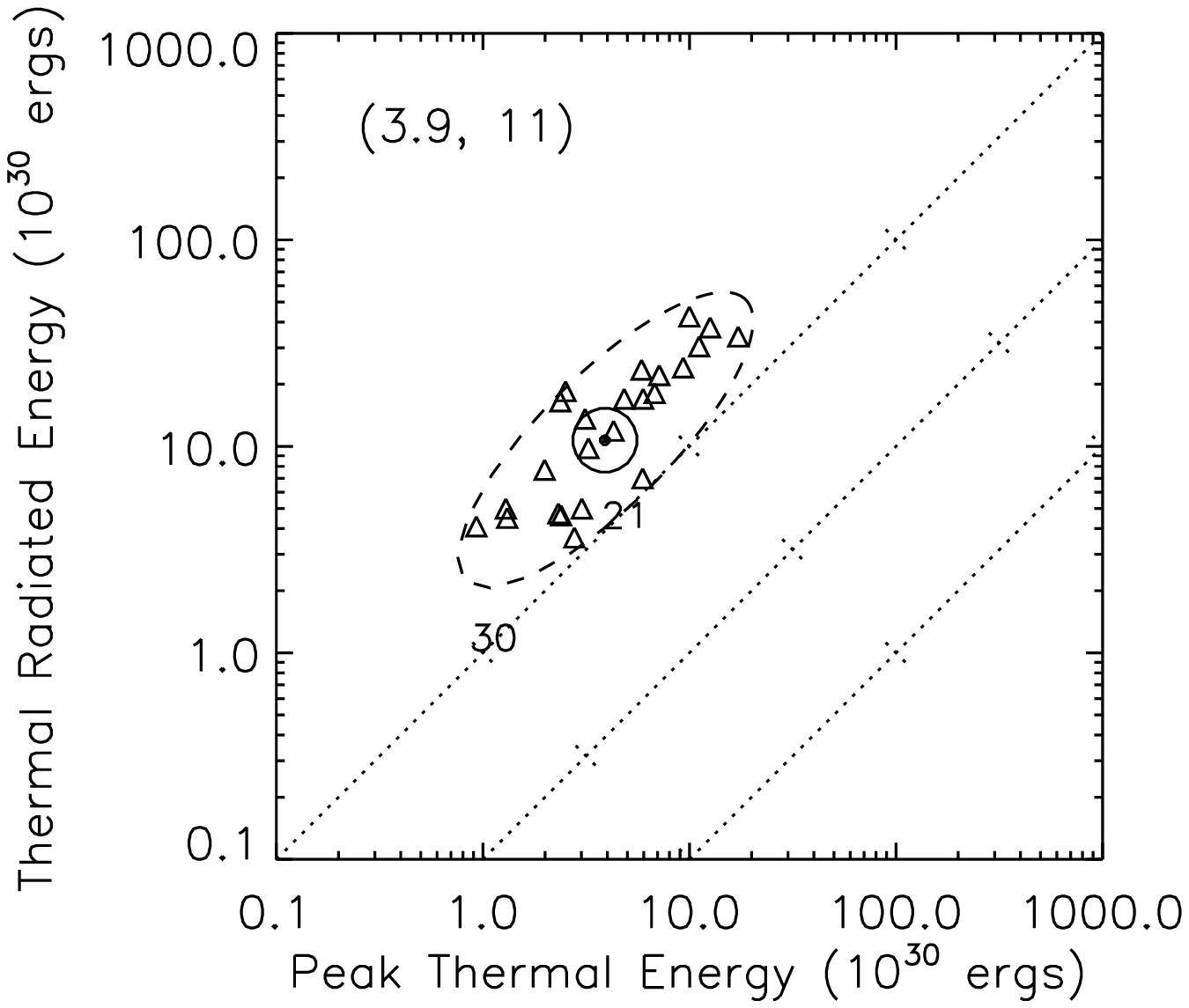}\label{1b}}
    \subfigure[][Peak thermal energy in the SXR-emitting plasma vs.\ energy in flare-accelerated nonthermal particles (electrons plus ions when available).]{\includegraphics[width=0.40\textwidth, bb=0 0 400 360]{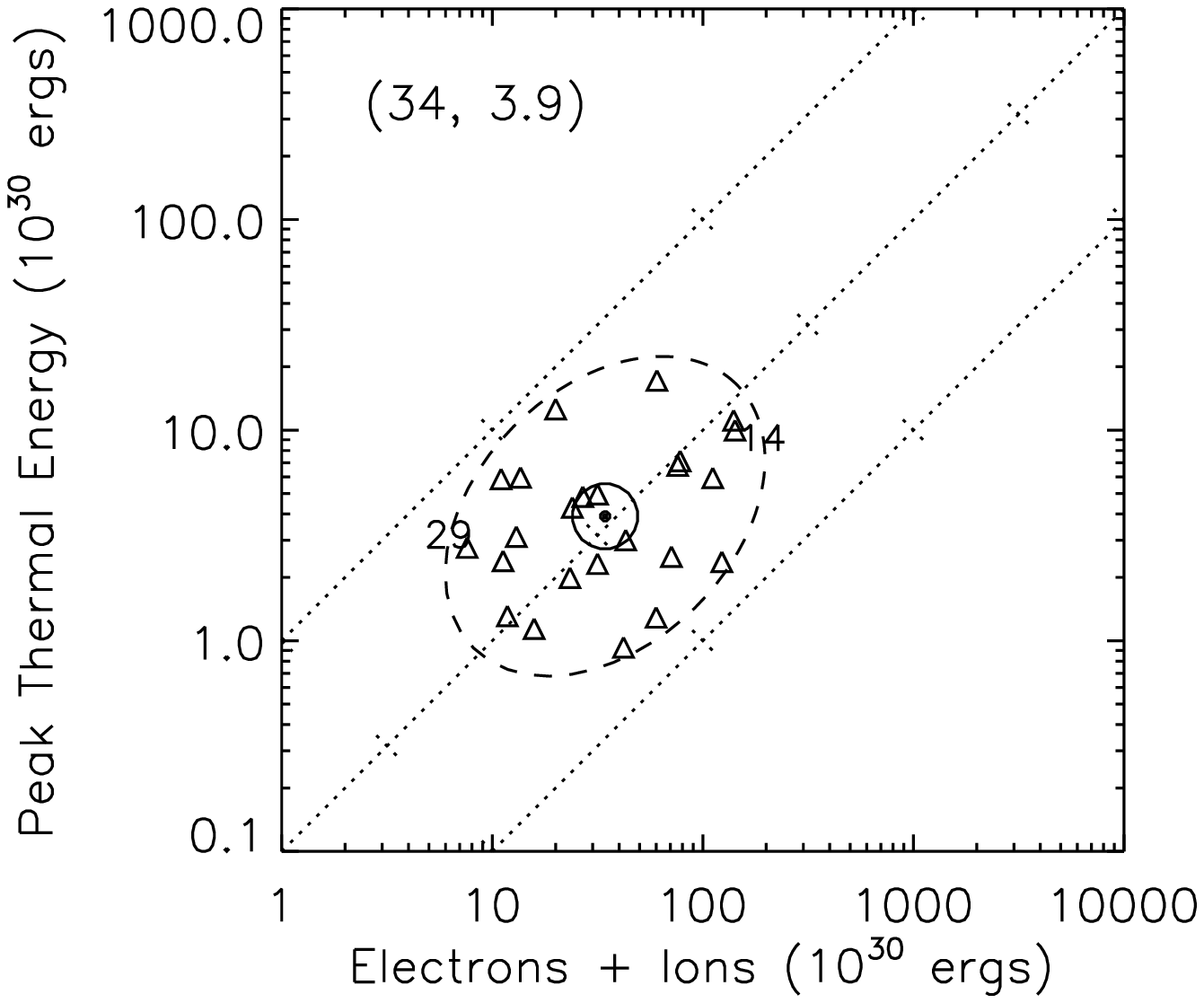}\label{1c}} \quad
    \subfigure[][Thermal radiated energy from the SXR-emitting plasma vs.\ the energy in flare-accelerated nonthermal particles (electrons plus ions when available).]{\includegraphics[width=0.40\textwidth, bb=0 0 400 360]{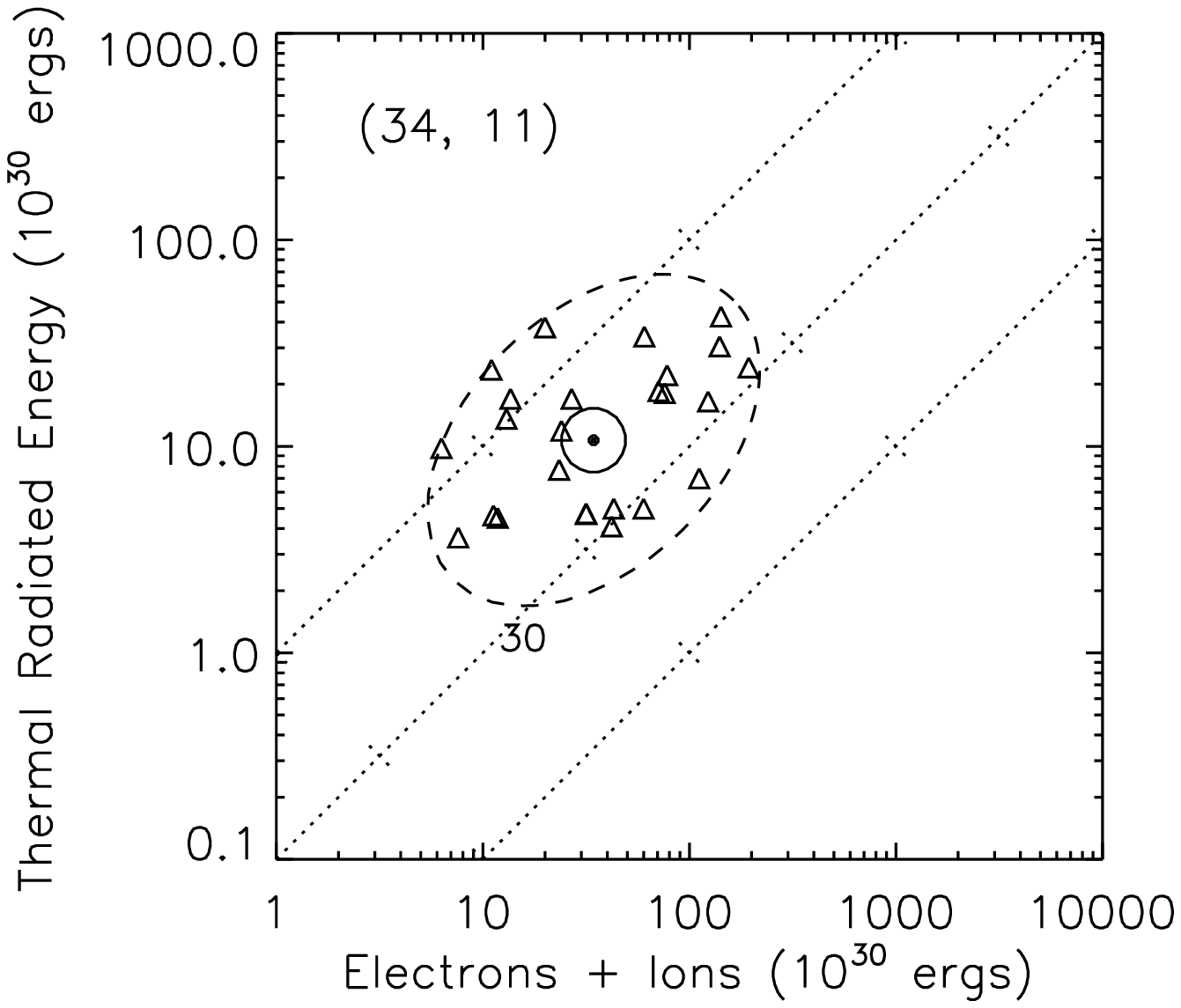}\label{1d}}
\end{center}
    \caption[]{\small{Scatter plots of different energy components, in units of $10^{30}$~ergs. Each plot includes all events for which measurements are available for both components. The points are indicated by triangles, except for the ``outlier'' points lying outside the $2\sigma$ ellipse (see below), which are instead labeled by their event numbers (Table~\ref{tbl-1}) and are located at the center of the respective numbers.  The location of the logarithmic centroid, defined by Equation~(\ref{eq_centroid}), for all the events in the plot is shown by a bullseye with its $X$ and $Y$ coordinates listed in the upper left corner of the plot.  The three diagonal dashed lines represent the 1\%, 10\% and 100\% ratios between the plotted components. Lines of constant logarithmic average event energy are shown by dashes every order of magnitude along the lines of constant ratio. The major and minor axes of the ellipse are defined by $\pm2$~times the RMS deviation of the points respectively parallel (Equation~(\ref{eq_RMSparal})) and perpendicular (Equation~(\ref{eq_RMSperp})) to the line of constant ratio passing through the centroid (see text and Appendix for discussion). Points outside this ellipse are considered as outliers and are discussed in Section \ref{outliers}}.
}
    \label{goes_thermal}
\end{figure}

\begin{figure}[pht]
\begin{center}
    \subfigure[][Energy in flare-accelerated ions vs.\ energy in flare-accelerated electrons.]{\includegraphics*[width=0.40\textwidth, bb=0 0 400 360]{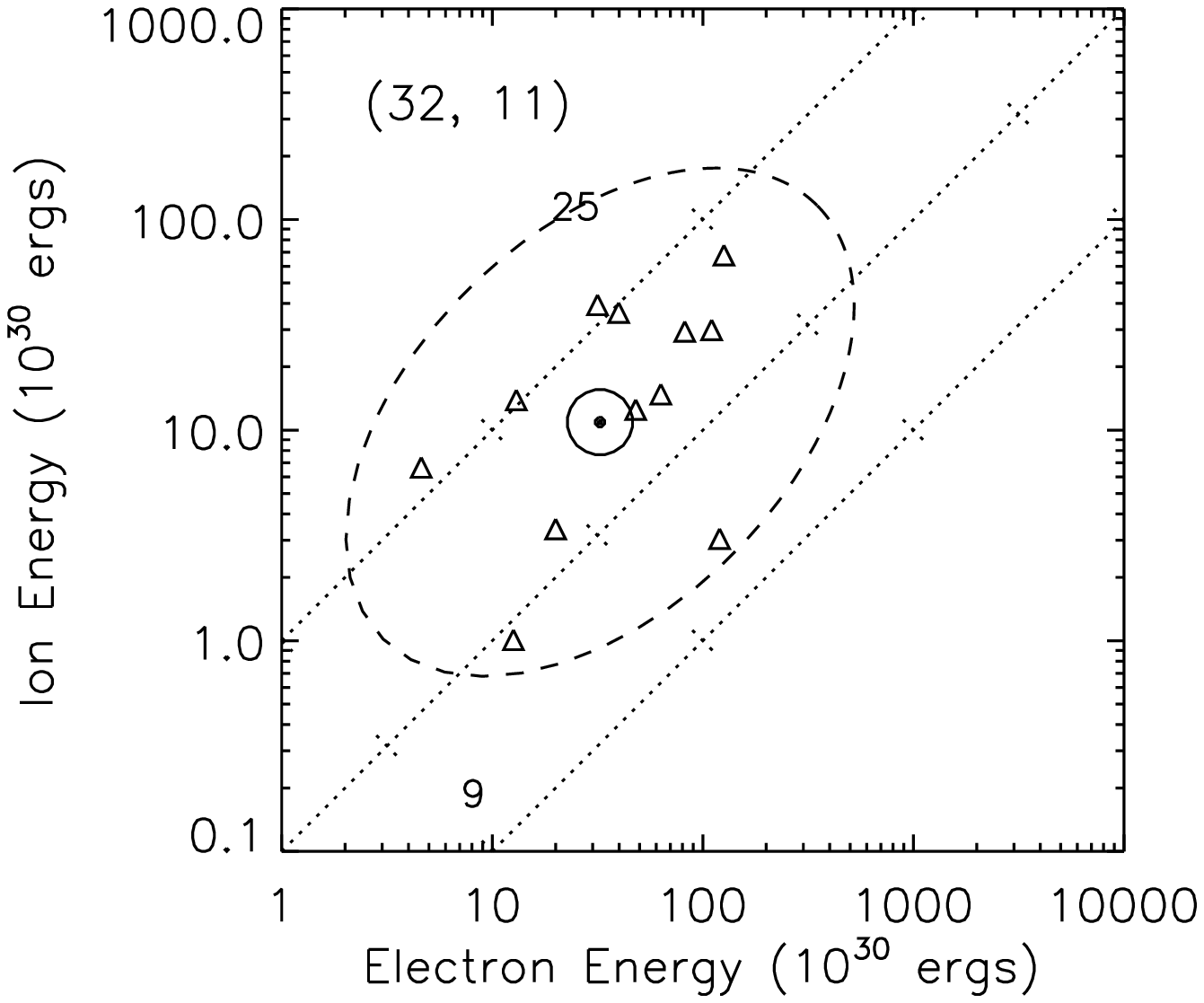}\label{2a}} \quad
    \subfigure[][Energy in SEPs vs.\ CME kinetic energy in the rest frame of the solar wind.]{\includegraphics[width=0.40\textwidth, bb=0 0 400 360]{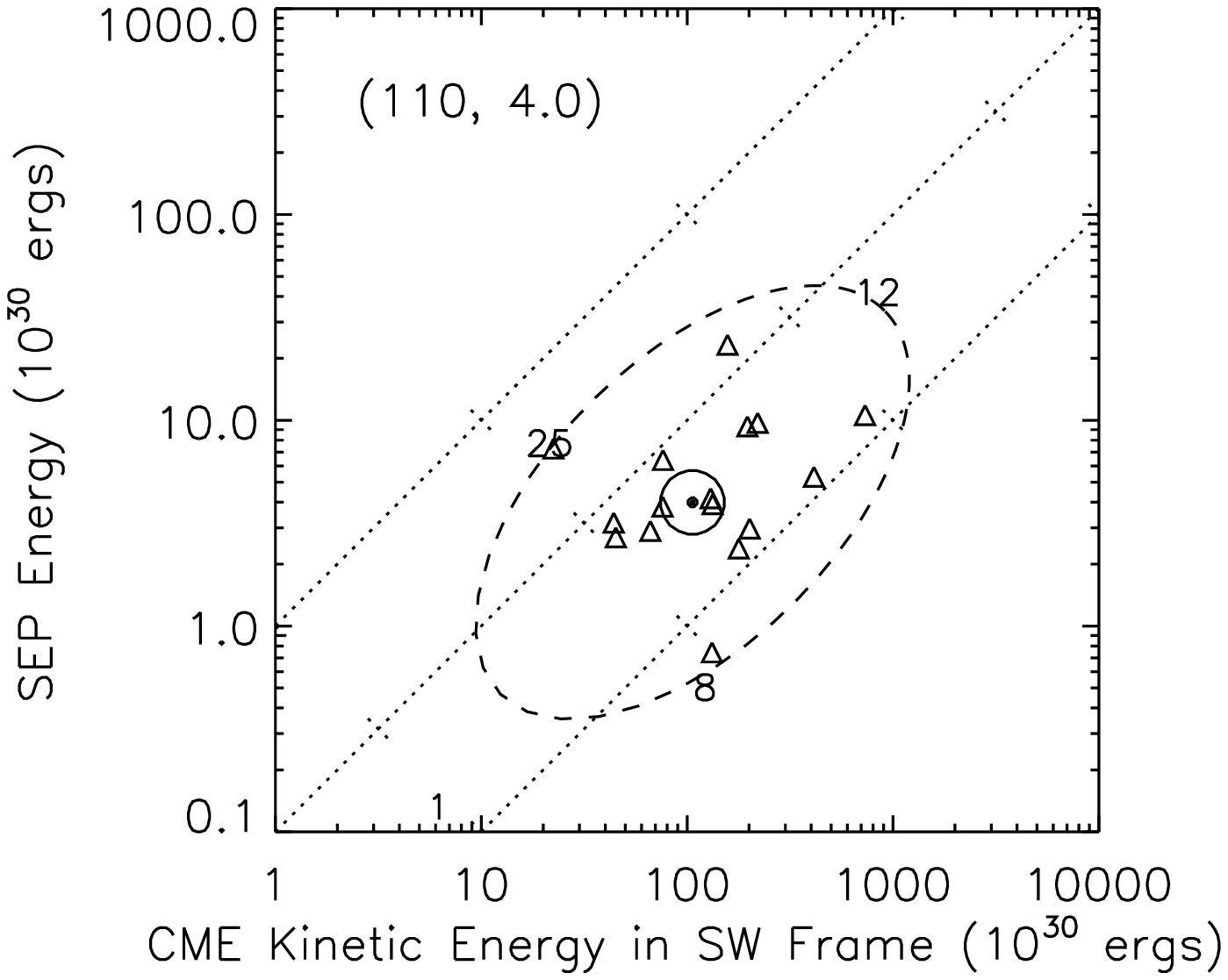}\label{2b}}
    \subfigure[][Energy in SEPs vs.\ energy in flare-accelerated ions.]{\includegraphics[width=0.40\textwidth, bb=0 0 400 360]{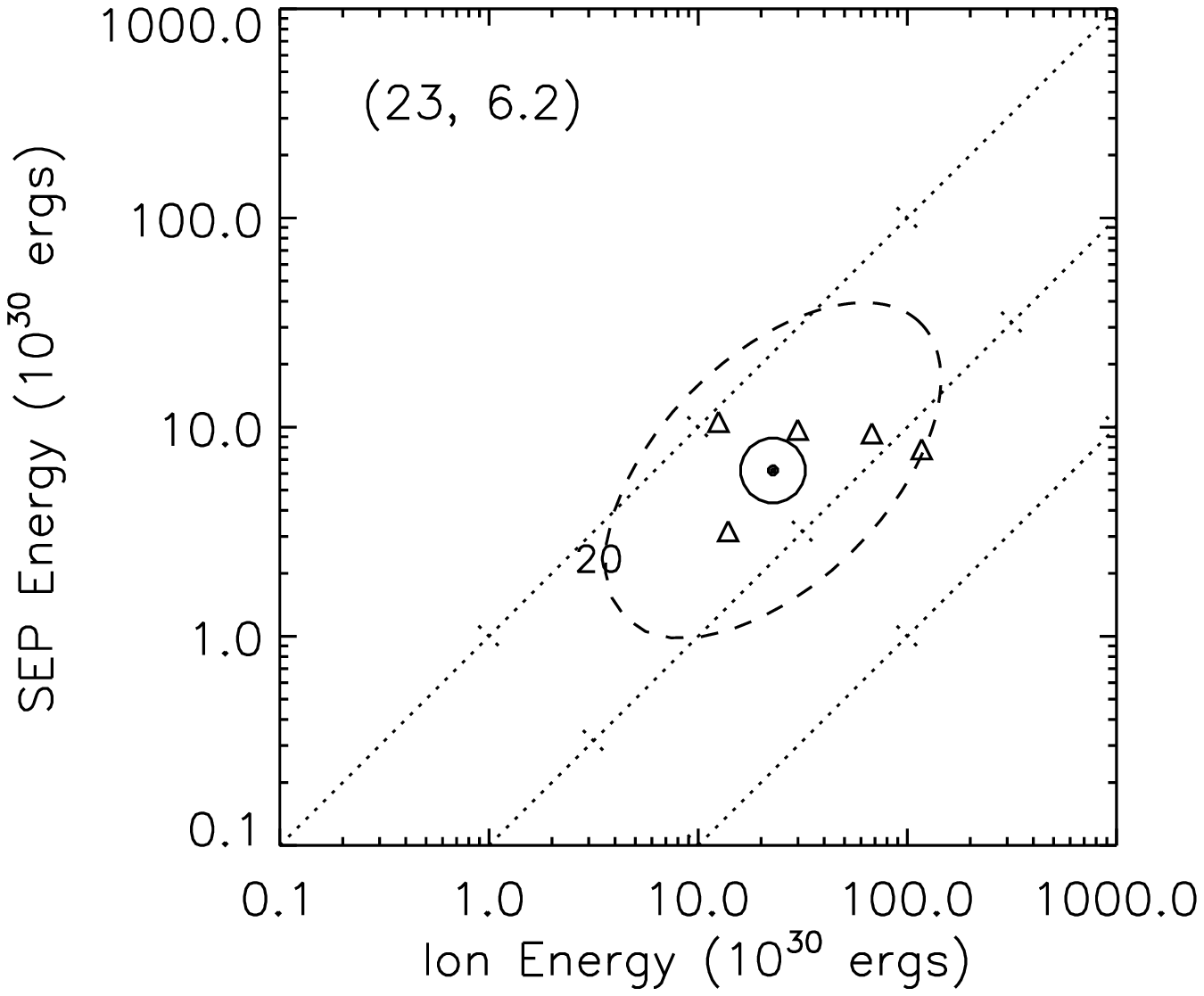}\label{2c}} \quad
    \subfigure[][Bolometric radiated energy vs.\ nonpotential magnetic energy in the \mbox{active} \mbox{region.}]{\includegraphics[width=0.40\textwidth, bb=0 0 400 360]{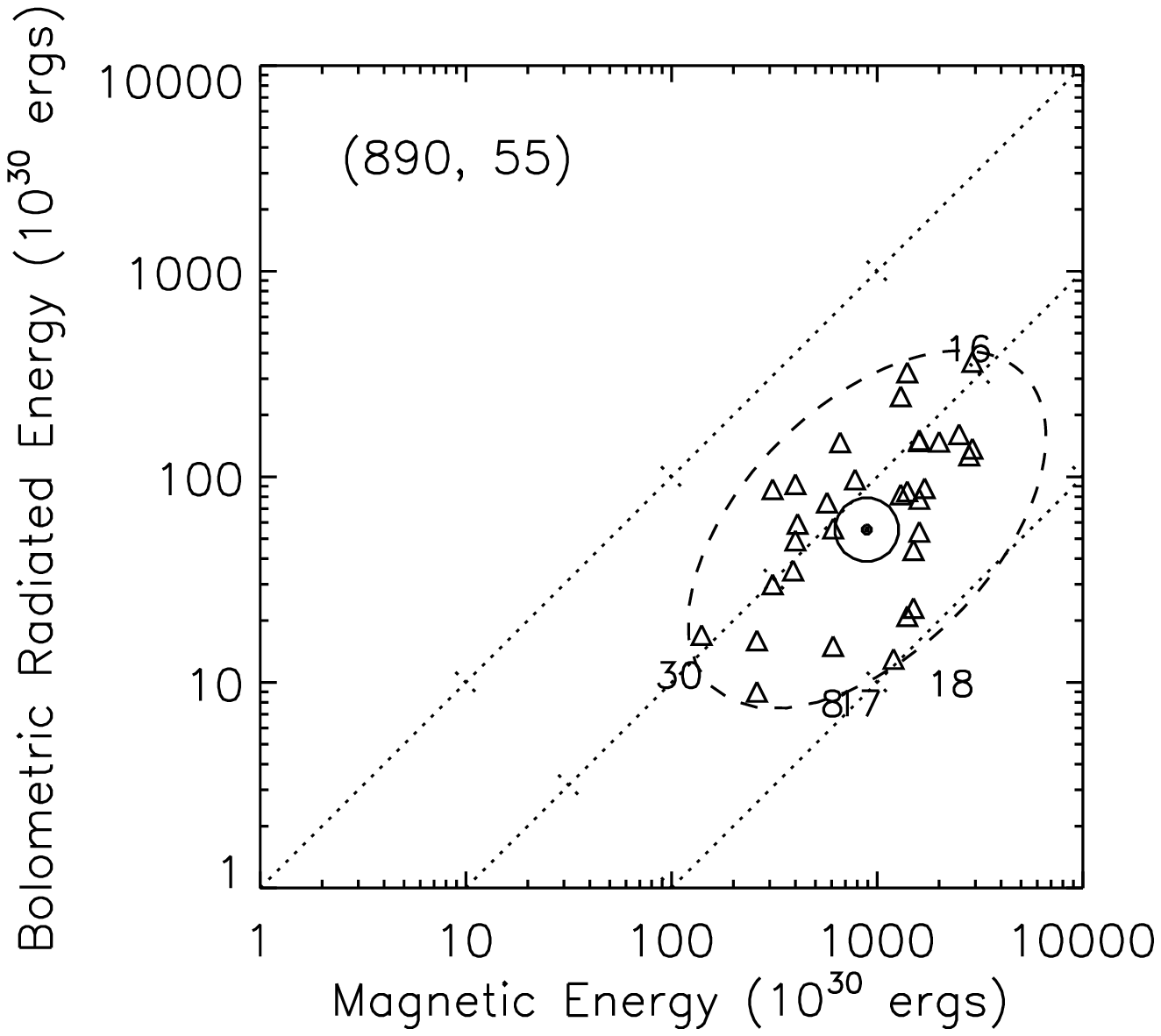}\label{2d}}
\end{center}
    \caption{Same as Figure~\ref{goes_thermal} for different combinations of energy components, as indicated on the axis labels.}
    \label{ions_electrons}
\end{figure}

\begin{figure}[pht]
\begin{center}
    \subfigure[][CME total energy (kinetic + potential, in the rest frame of the Sun) vs.\ the free (nonpotential) magnetic energy of the \mbox{active} \mbox{region.}]{\includegraphics[width=0.40\textwidth, bb=0 0 400 360]{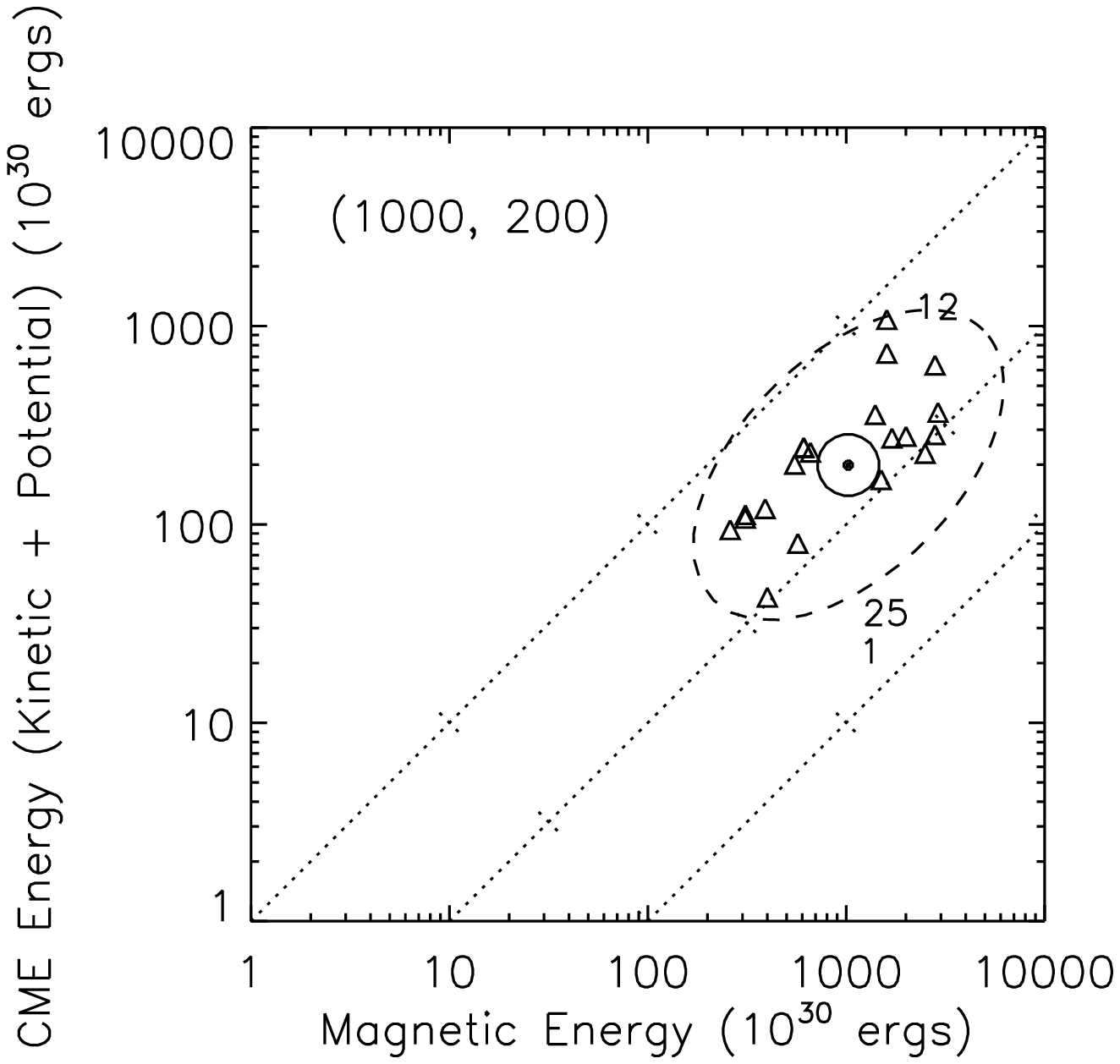}\label{3a}} \quad
    \subfigure[][Bolometric radiated energy vs.\ the CME total energy (kinetic + potential, in the rest frame of the Sun).]{\includegraphics[width=0.40\textwidth, bb=0 0 400 360]{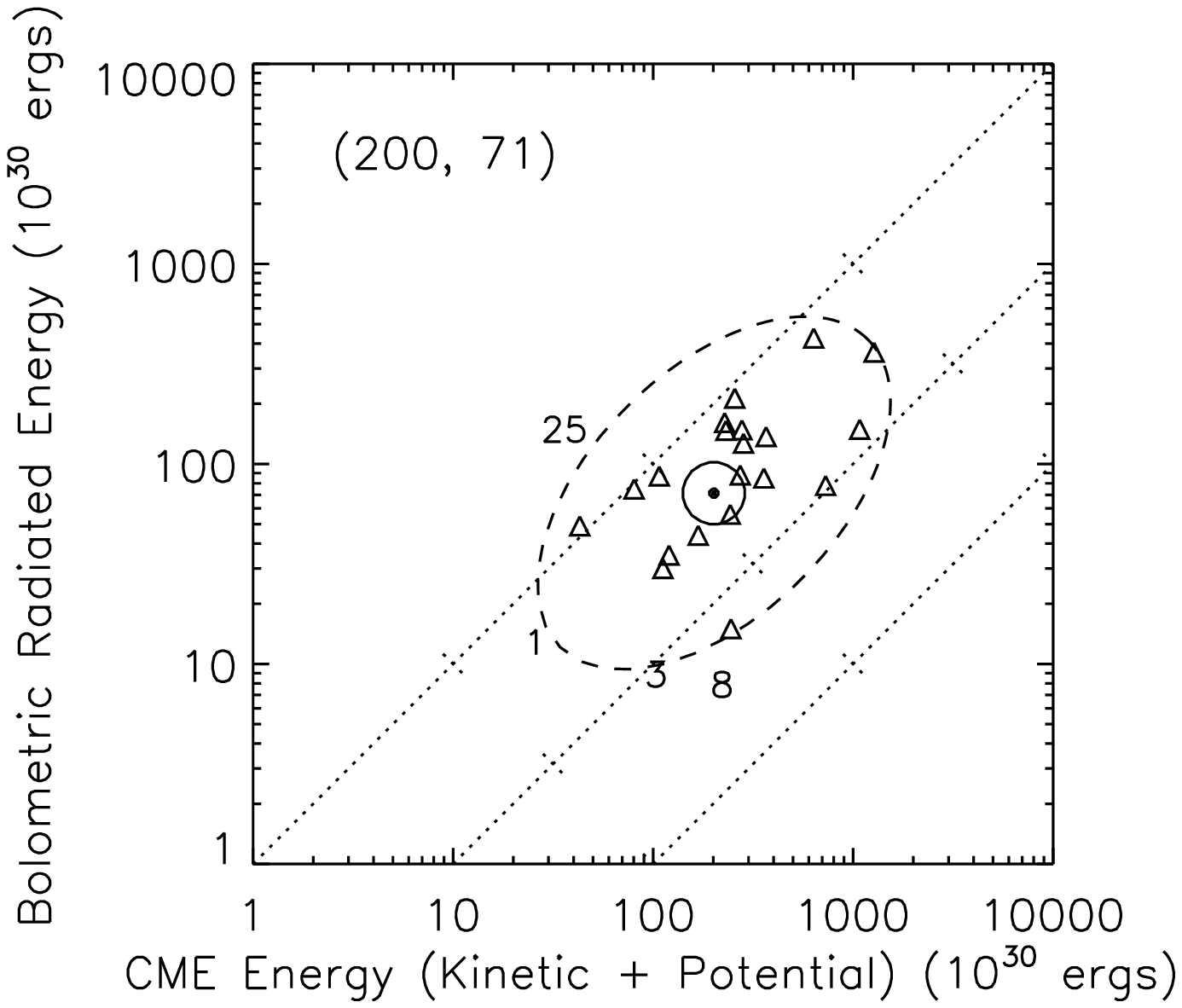}\label{3b}}
    \subfigure[][Energy in flare-accelerated nonthermal particles (\mbox{electrons} and ions) vs.\ bolometric radiated energy.]{\includegraphics[width=0.40\textwidth, bb=0 0 400 360]{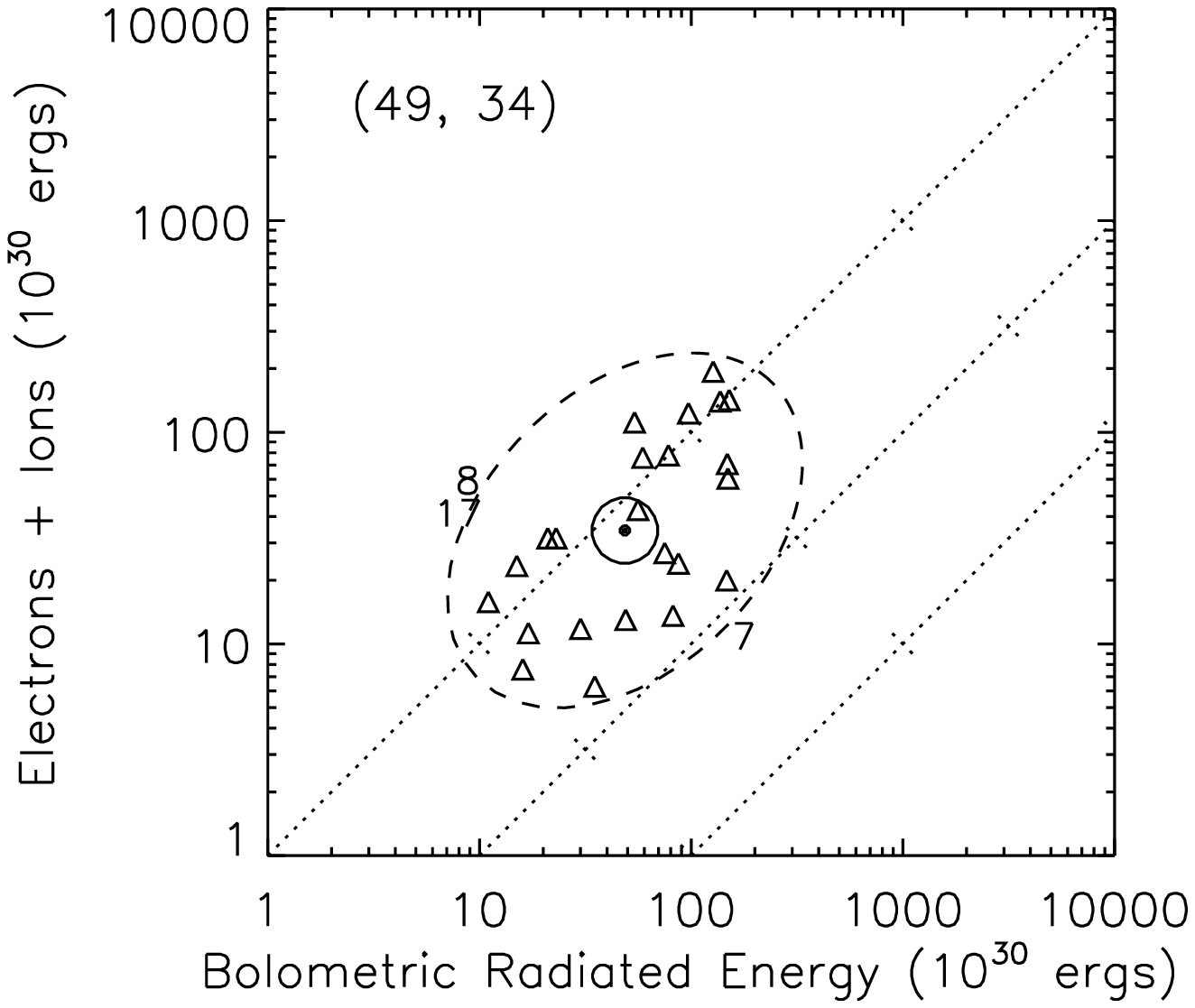}\label{3c}} \quad
     \subfigure[][Total energy radiated by SXR-emitting plasma vs.\ bolometric radiated energy.]{\includegraphics[width=0.40\textwidth, bb=0 0 400 360]{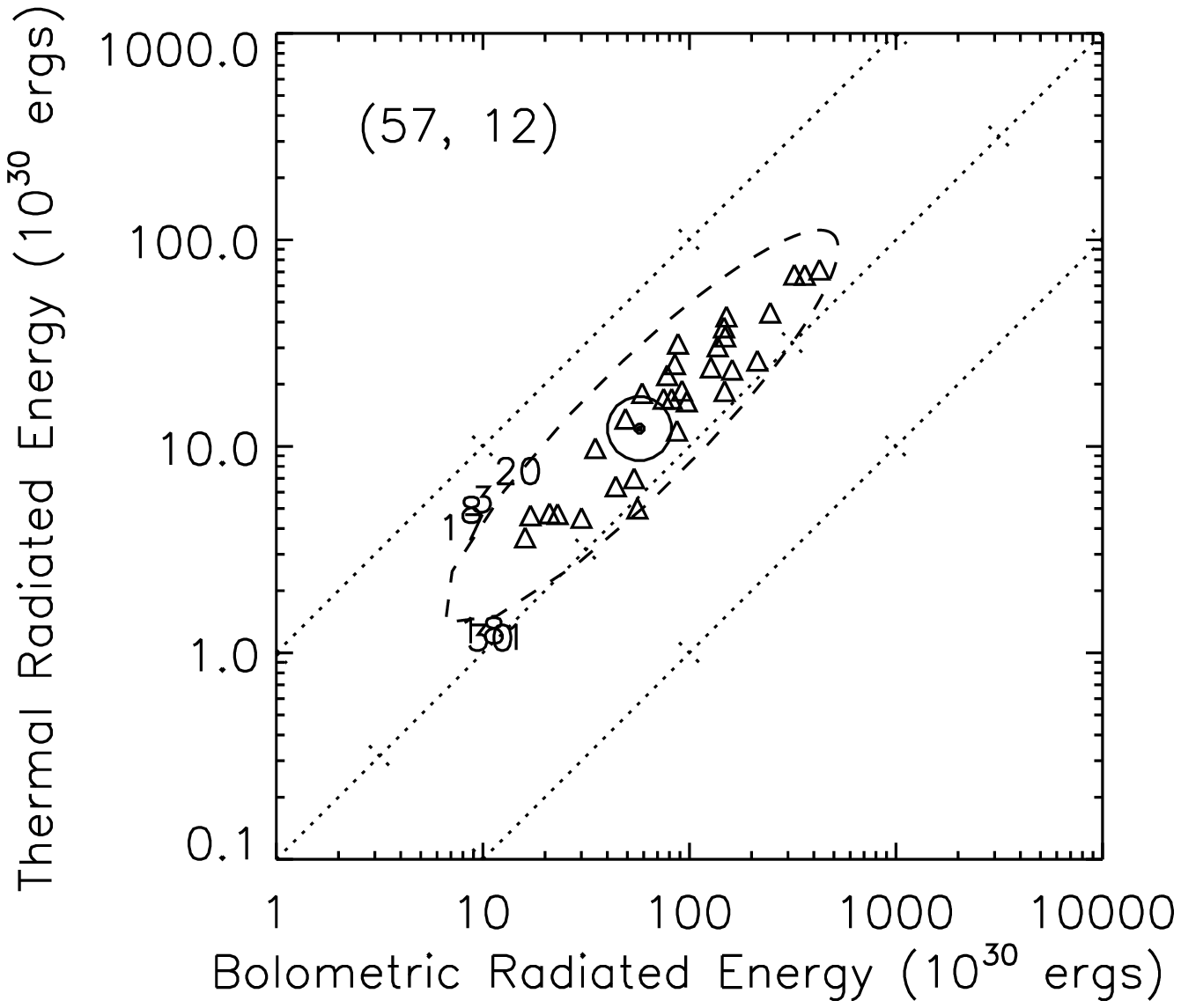}\label{3d}}

\end{center}
    \caption{Same as Figure~\ref{goes_thermal} for different combinations of energy components, as indicated on the axis labels.}
    \label{cme_mag}
\end{figure}

The component energy comparisons are discussed in the following subsections, with associated plots given in Figures \ref{goes_thermal} to \ref{cme_mag}.  All plots have the same four-order-of-magnitude range on each axis, so that the degree of spread in a particular energetic component can be readily visualized.  In each plot, all events that have measured energies for both components are shown.  The points are indicated by triangles, except for the ``outlier'' points lying outside the $2\sigma$ ellipse (see below), which are instead labeled by their event numbers (Table~\ref{tbl-1}) and are located at the center of the respective numbers. Events with only upper and/or lower limits are generally not shown.  However, we have included pertinent data for the behind-the-limb Event \#5 and for Event \#25, for which there is some ambiguity in the CME kinetic energy and we have hence used the geometric mean of the two estimates. The logarithmic centroid is shown by a bullseye with its $X$ and $Y$ coordinates, calculated using Equation (\ref{eq_centroid}), given in the upper left corner of the plot.  The three diagonal dotted lines are lines of constant ratio $R$ as defined by Equation (\ref{eq_R}), with $R = Y/X = 100$\%, 10\% and 1\%.  These lines each have tick marks showing the overall ``size'' of the event $A = \sqrt{XY}$ -- see Equation~(\ref{eq_A}).  The dashed-line ellipse shows the $\pm2\sigma$ locus; the widths of this ellipse perpendicular (Equation~(\ref{eq_RMSperp})) and parallel (Equation~(\ref{eq_RMSparal})) to the lines of constant ratio are measures of the $2\sigma$ spread in the energy ratio, $R$, and the event size, $A$, respectively. Points outside this ellipse are considered as ``outliers'' and will be discussed in Section~\ref{outliers}.

Table~\ref{tbl-scatterparams} lists, for each plot, the energetic components involved, the value of the logarithmic centroid energies and their ratio, the root-mean-square (RMS) spreads in the values of the ratio $R$ and the size $A$.  Also, to quantify possible trends of one parameter vs.\ the other, Table~\ref{tbl-scatterparams} lists the Spearman's rank correlation coefficient $\rho$, a quantity that measures the correlation between their rank orders (lowest $\rightarrow$ highest) of the variables. The formal equations used to determine these different parameters are given in Appendix~\ref{appendix}.

\begin{deluxetable}{cllccccccc}

    \tabletypesize{\scriptsize}
    \tablewidth{0pt}
    \tablecaption{Parameters represented in scatter plots.}
    \label{tbl-scatterparams}

\tablehead{\colhead{Fig.}      & \multicolumn{2}{c}{Plotted Components} & \multicolumn{2}{c}{Log. Centroid\tablenotemark{*}}& $R$\tablenotemark{**}   &  \multicolumn{2}{c}{ RMS\tablenotemark{\dagger} } & $\rho$\tablenotemark{\ddagger} & No. of \\

 \colhead{No. } & {$X$-axis} & {$Y$-axis} & \colhead{$X$} & \colhead{$Y$} &  &   \colhead{$R$ } & \colhead{$A$ }&& Events \\

 &           &             & \colhead{$10^{30}$ ergs}&\colhead{$10^{30}$ ergs}&   & & }
\startdata

    1a    & Rad. from Hot Plasma & GOES 1--8 $\mbox{\AA}$        & 12   & 0.6 & 0.05      & 0.17      & 0.51 & 0.96 & 38 \\
    1b    & Peak Thermal Energy  & Rad. from Hot Plasma          & 3.9  & 11  & 2.7       & 0.22      & 0.34 & 0.82 & 26 \\
    1c    & Electrons+Ions       & Peak Thermal Energy           & 34   & 3.9 & 0.11      & 0.43      & 0.31 & 0.36 & 26 \\
    1d    & Electrons+Ions       & Rad. from Hot Plasma          & 34   & 11  & 0.31      & 0.43      & 0.34 & 0.46 & 26 \\
    2a    & Electrons            & Ions                          & 32   & 11  & 0.34      & 0.63      & 0.52 & 0.45 & 14 \\
    2b    & CME KE (SW frame)    & SEP                           & 110  & 4.0 & 0.04      & 0.49      & 0.47 & 0.47 & 20 \\
    2c    & Ions                 & SEP                           & 23   & 6.2 & 0.27      & 0.38      & 0.35 & 0.20 &  6 \\
    2d    & Magnetic             & Bolometric                    & 890  & 55  & 0.06      & 0.43      & 0.38 & 0.56 & 37 \\
    3a    & Magnetic             & CME KE+PE                     & 1000 & 200 & 0.19      & 0.39      & 0.34 & 0.68 & 23 \\
    3b    & CME KE+PE            & Bolometric                    & 200  & 71  & 0.35      & 0.43      & 0.38 & 0.54 & 24 \\
    3c    & Bolometric           & Electrons+Ions                & 49   & 34  & 0.71      & 0.47      & 0.35 & 0.37 & 26 \\
    3d    & Bolometric           & Rad. from Hot Plasma          & 57   & 12  & 0.21      & 0.21      & 0.47 & 0.92 & 38 \\

\enddata

\tablenotetext{*}{$X$ and $Y$ values of the logarithmic centroid, computed using Equation~(\ref{eq_centroid}).}
\tablenotetext{**}{$R = Y/X$, the ratio of $Y$ and $X$ values of the logarithmic centroid computed using Equation~(\ref{eq_R}).}
\tablenotetext{\dagger}{RMS (root mean square) values of $R = Y/X$ and $A = \sqrt{XY}$, computed using Equations~(\ref{eq_RMSperp}) and~(\ref{eq_RMSparal}). The RMS values of $R$ and $A$, respectively, quantify the scatter perpendicular and parallel to the line of constant energy ratio that passes through the logarithmic centroid.}
\tablenotetext{\ddagger}{Spearman's rank correlation coefficient -- a non-parametric measure of statistical dependence between two variables -- see Equation~(\ref{spearman}).}

\end{deluxetable}

\subsection{Radiated Energy in the GOES 1 -- 8 $\mbox{\AA}$ Band vs.\ Radiated Energy from SXR-emitting Plasma}

Figure~\ref{goes_thermal}a shows the scatter plot for the radiated energy in the GOES 1~--~8~$\mbox{\AA}$ band vs.\ the total energy radiated from the hot SXR-emitting plasma.  The points are closely rank-correlated ($\rho = 0.96$) and also cluster very closely in the perpendicular ($R$) direction, showing that the energy radiated in the GOES 1 -- 8 $\mbox{\AA}$ band is a relatively constant fraction ($R=0.05$) of the total energy radiated from the SXR-emitting plasma.  Indeed, a regression analysis of the data in Table~\ref{tbl-1} shows that the ratio (best estimate $\pm1\sigma$) of the total energy radiated by SXR-emitting thermal plasma in the flare to the observed GOES 1~--~8~$\mbox{\AA}$ flux is $15.4 \pm 0.8$. This strong correlation is not surprising since both plotted energy components are calculated from the GOES SXR fluxes; the scatter about the trend line arises from the differences in the temperatures of the different events.

\subsection{Thermal Radiated Energy vs.\ Peak Thermal Energy}

Figure~\ref{goes_thermal}b shows the scatter plot of the total energy radiated from hot SXR-emitting plasma vs.\ the peak thermal energy content of that plasma. The relatively tight correlation ($\rho = 0.82$) between these two components is expected, since both parameters refer to the same SXR-emitting plasma. Event \#30 (the M6.4 event on 2005~August~25) is the most extreme outlier but it is almost equally weak in both energy components. On average, the total energy radiated exceeds the peak thermal energy content by a factor of $\sim$3 ($R = 2.7$ in Table \ref{tbl-scatterparams}), implying continuous re-energization of the SXR-emitting material as the flare progresses.

\subsection{Peak Thermal Energy vs.\ Energy in Flare-Accelerated Nonthermal Particles}

Figure~\ref{goes_thermal}c shows the scatter plot for the peak thermal energy in the hot SXR-emitting plasma vs.\ the energy in flare-accelerated nonthermal particles (electrons plus ions, when available). There is substantially greater spread in the points compared to Figure~\ref{goes_thermal}b,
but nevertheless a reasonable bunching of the points. The maximum spread of the points is less than two orders of magnitude in either parameter. On average, the energy in the flare-accelerated non-thermal particles exceeds the peak thermal energy by almost an order of magnitude ($R = 0.11$), indicating that there is easily sufficient power in the particles to create the SXR-emitting thermal plasma. This conclusion is reinforced by the fact that the energy in flare-accelerated electrons is a lower limit (see Section~\ref{electrons}) and is in agreement with earlier comparisons of flare-accelerated electrons versus peak thermal energy -- see, e.g., \cite{1986NASACP...2439..505D} and \cite{2002SoPh..210..287S}.

\subsection{Thermal Radiated Energy vs.\ Energy in Flare-Accelerated Nonthermal Particles}

Figure~\ref{goes_thermal}d shows the scatter plot for the total energy radiated by the SXR-emitting thermal plasma vs.\ the energy in flare-accelerated nonthermal particles (electrons and ions when available.  This figure combines information already evident in Figure~\ref{goes_thermal}b and Figure~\ref{goes_thermal}c.
It shows that the energy in accelerated electrons and ions during a flare is not only sufficient to supply the peak energy of the SXR-emitting plasma (Figure~\ref{goes_thermal}c),
but it is also high enough (by a factor of $\sim 3$, $R = 0.31$) to account for the radiation from this plasma throughout the event.  As discussed in Section~\ref{SEEstudied}, it follows that a significant fraction of the energy in flare-accelerated nonthermal particles is deposited by thermal conduction into lower-temperature plasma and ultimately radiated in optical and EUV wavebands \citep[see][]{2005JGRA..11011103E}.  Again, Event \#30 (the M6.4 event on 2005 August 25) is the only ``outlier'' in this plot, reflecting the low values of both the thermal and nonthermal energy components.

\subsection{Flare-Accelerated Ions vs.\ Electrons}

Figure~\ref{ions_electrons}a shows the scatter plot for the energy in flare-accelerated ions, as determined from the {\em RHESSI} gamma-ray observations (Section~\ref{ions}), vs.\ the energy in flare-accelerated electrons, as determined from {\em RHESSI} hard X-ray observations (Section~\ref{electrons}).

As pointed out in Section \ref{electrons}, the energy in electrons is critically dependent on the low-energy cutoff, E$_{\rm min}$, that is assumed for the electron spectrum. Since the largest value of E$_{min}$ that gives an acceptable fit to the data is used for each spectrum, the total electron energy values are lower limits with order-of-magnitude uncertainties. As explained in Section~\ref{ions}, the situation for the ion energies is even worse, both because of the spread in the observed 2.223~MeV fluences and because of the need to extrapolate the ion flux at energies above 30~MeV, as derived from these 2.223~MeV line fluences, to the ion flux above 1~MeV. Because of these large uncertainties in both the electron and ion energies, there is a much wider scatter than in the plots in Figure~\ref{goes_thermal}.  However, with some notable exceptions that have almost two orders of magnitude more energy in the electrons than in the ions (Events \#9 and \#15), the electron and ion energies are generally comparable within an order of magnitude. This result is in agreement with the claims by \cite{1995ApJ...455L.193R} and \cite{2000IAUS..195..123R} and has significant consequences for particle acceleration models.

\subsection{SEP Energy vs.\ CME Kinetic Energy in the Solar Wind Rest Frame}
\label{SEPvsCME_KESW}
Figure~\ref{ions_electrons}b shows the scatter plot for the energy in the accelerated SEP population vs.\ the kinetic energy of the CME in the rest frame of the solar wind. We use the solar wind rest frame since a shock can be formed and SEPs accelerated only if the CME is traveling at least as fast as the solar wind speed \citep{2008SSRv..136..285M}. Lacking knowledge of the solar wind speed low in the corona for each event, we have simply subtracted 400 km~s$^{-1}$ from the measured CME speed in order to estimate the kinetic energy available for accelerating particles via shock acceleration.

Most of the SEP values cluster between 1\% and 10\% of the CME kinetic energy. Comparing the nine events that are common to both \citet{2008AIPC.1039..111M} and this study (Events \#2, 4, 7, 8, 11, 12, 13, 14, and 16), the SEP/CME ratio was 5.8\% in \citet{2008AIPC.1039..111M} and is 4\% here ($R = 0.04$ in Table~\ref{tbl-scatterparams}). Overall, as a result of several changes in the analysis, the SEP energy estimates in this paper are reduced from those in \citet{2005JGRA..11009S18M,2008AIPC.1039..111M} by an average of ~40\%. One of these changes is the adoption of the longitude correction of \cite{2006ApJ...653.1531L}, which
results in changes of as much as a factor of two in the energy estimates of individual SEP events.  For the nine events in common with this paper and \citet{2008AIPC.1039..111M}, the differences due to longitude/latitude corrections alone ranged from -52\% to +51\% with a mean difference of -16\%; for all 20 events shown in Figure~\ref{ions_electrons}b the average effect is a $\sim$10\% decrease in energy.  Another change is the adoption of new corrections for particles crossing 1~AU multiple times and for adiabatic energy loss that are based on the energy and species-dependent simulations of \citet{2010JGRA..11506101C}; these changes resulted in a further decrease averaging $\sim$30\% in the SEP energy content of the nine events in common with \citet{2008AIPC.1039..111M}.  Finally, the SEP spectra in this study were integrated from 0.03~MeV~nucleon$^{-1}$ to 300~MeV~nucleon$^{-1}$, rather than from 0.01~MeV~nucleon$^{-1}$ to 1000~MeV~nucleon$^{-1}$ as in \citet{2004JGRA..10910104E} and \citet{2005JGRA..11009S18M,2008AIPC.1039..111M}.  The increase in the low-energy limit to 0.03~MeV~nucleon$^{-1}$ represents a more realistic threshold for injection into the shock acceleration process \citep[see, e.g.,][]{2012AdSpR..49.1067L}, and results in a typical reduction in the energy content by $\sim$5\%.  The change in the upper limit has a negligible effect.

Overall, this new analysis confirms that the SEP energy is a small, but not insignificant, fraction of the CME kinetic energy in most large events.

\subsection{SEP Energy vs.\ Energy in Flare-Accelerated Ions}

Figure~\ref{ions_electrons}c shows the scatter plot for the total energy in SEPs vs.\ the energy in flare-accelerated ions, as determined from {\em RHESSI} gamma-ray observations. There are only a limited number of events that can be compared but in those few cases there is comparable energy in the ions and SEPs ($R = 0.27$).  At first sight, this appears to conflict with the study by Mewaldt (2012, in preparation) that shows the number of $>$30 MeV SEP protons measured in interplanetary space is generally much higher than the number of $>$30~MeV protons interacting in the solar atmosphere \citep{2009ApJ...698L.152S}.  However, this difference can be accounted for in a number of ways.  First, as noted in Section~\ref{ions}, the inferred energy in flare-accelerated ions is based on a very uncertain spectral extrapolation over more than an order of magnitude in proton energy (from 30~MeV down to 1~MeV).  To obtain the energy in flare-accelerated ions (Section~\ref{ions}), we have assumed a spectral index of 4, which results in a significantly higher total energy than would be obtained if the extrapolation was performed with the much lower spectral indices representative of SEP spectra measured in situ at 1~AU.  Further, for the well-observed 2003~October~28 flare (Event \#12), we obtained an ion spectral index of 3.4 at energies between $\sim$3 and 50~MeV, so that the energy content in ions could be significantly lower than we have used here, especially if the spectrum hardens even more between, say, 1~and 20~MeV, a feature that is clearly seen in SEP spectra.  Second, SEP protons typically carry $\gapprox $80\% of the SEP ion energy, whereas flare-accelerated protons carry only about one-third of the ion energy.

\subsection{Bolometric Radiated Energy vs.\ Magnetic Energy}\label{bol_mag_sec}

Figure~\ref{ions_electrons}d shows the scatter plot for the bolometric radiated energy  vs.\ the nonpotential (free) magnetic energy in the active region.  Here we have a relatively tight bunching with little more than 1.5 orders of magnitude range in each parameter. It should be emphasized that the plotted bolometric radiated energies are, with the exception of the five events noted in Table~\ref{tbl-1}, not directly measured but rather estimates made using the FISM model \citep{2007SpWea...507005C, 2008SpWea...605001C}, and the magnetic energy is only good to an order of magnitude. With this proviso, we find that the average bolometric radiated energy is $\sim$6\% of the free magnetic energy ($R = 0.06$), and in all cases the available magnetic energy exceeds the bolometric radiated energy by at least half an order of magnitude. This is consistent with the well-accepted notion that the reservoir of magnetic energy is sufficient to power the main components of the flare.

\subsection{CME Energy vs.\ Magnetic Energy}

Figure~\ref{cme_mag}a shows the scatter plot for the CME total energy (potential + kinetic) in the rest frame of the Sun vs.\ the nonpotential energy in the magnetic field.  The CME energy is, on average, only a small fraction ($R = 0.19$) of the magnetic energy, similar to the value $R=0.06$ found for the ratio of bolometric radiated energy to magnetic energy (Section~\ref{bol_mag_sec}; Figure~\ref{ions_electrons}d).  While bearing in mind the very approximate values of the latter, it nevertheless appears, from the results of this and the previous subsection, that much of the available magnetic energy (some two-thirds) is retained in the active region (i.e., the field does not return to a fully potential state), even after the flare and the ejection of the CME.  This result is consistent with the generous limits established by \citet{2012ApJ...750...24M} on the possible free energy that an active region magnetic field can hold before it erupts.

\subsection{Bolometric Radiated Energy vs.\ CME Energy }

Figure~\ref{ions_electrons}d and~\ref{cme_mag}a show, respectively, that the available magnetic energy is about 15 times the bolometrically radiated energy in the flare ($R = 0.06$) and about 5 times the CME energy ($R = 0.19$). Given the substantial overlap of events common to both plots (Tables~\ref{tbl-1} and~\ref{tbl-scatterparams}), this indicates that, on average, the energy in the CME is larger than the energy radiated by a factor of $\sim$3.  Figure~\ref{cme_mag}b confirms this result by showing the scatter plot for bolometric radiated energy vs.\ the CME total energy (kinetic + potential).  On average, the bolometric energy is indeed about half an order of magnitude less than the CME energy ($R = 0.35$).

\subsection{Flare-Accelerated Particle Energy vs.\ Bolometric Radiated Energy}

Figure~\ref{cme_mag}c shows the scatter plot for the total energy in flare-accelerated particles (electrons plus ions) vs.\ the bolometric radiated energy.  This figure shows that the energy in accelerated particles during a flare is comparable to the total bolometric radiated energy from the flare ($R = 0.71$), with the ratio being greater than unity in some events and less than unity in others.  It must be recalled that while the bolometric radiated energies are accurate to within a factor of $\sim$2 to 3, the energies in flare-accelerated particles are uncertain to at least an order of magnitude. The energies in electrons are most probably lower limits and may well underestimate the true energy content by up to an order of magnitude. The energies in ions may, however, be overestimates, depending on the power-law index used for the spectral extrapolation. We can tentatively conclude, however, that there is sufficient energy in the flare-accelerated particles to account for {\it all} the energy radiated in the flare. However, this conclusion must somehow be verified by more accurate estimates of the energy in the flare-accelerated electrons and ions.

\subsection{Energy Radiated by Soft X-Ray Emitting Plasma vs.\ Bolometric Energy}

Figure~\ref{cme_mag}d shows the scatter plot for the total energy radiated by SXR-emitting plasma vs.\ the bolometric radiated energy. The relatively tight correlation ($\rho = 0.92$) is to a large extent due to the fact that most of the bolometric radiant energies were computed using the FISM model, which uses the radiated energy from SXR-emitting plasma to estimate the bolometric energy.  Nevertheless, the data show that only about one-fifth of the bolometric energy radiated by a flare is radiated by the SXR-emitting plasma ($R=0.21$).

\section{Discussion}
\label{discussion}

The comparisons of the energetics of the different components shown in the scatter plots of Figures~\ref{goes_thermal}--\ref{cme_mag} are summarized in Table~\ref{tbl-scatterparams}.

In attempting to draw any definitive conclusions from these comparisons, we must keep in mind the limitations of our analysis. The events in our list cover the period from {\em RHESSI}'s launch in February~2002 through 2006.  They include nine of the eleven X5 or greater events that occurred in this time period.  They also include the largest SEP events and all significant {\em RHESSI} gamma-ray line events, and hence represent events where a significant number of ions were accelerated to high energies, either in the flare, at the CME shock, or at both locations. Most of the CMEs that we have included have kinetic energies of $\gapprox 10^{32}$~erg.  According to Figure~8 of \citet{2004JGRA..10912105G}, only 16 CMEs observed from 1997 -- 2002 had kinetic energy $>$1.4$ \times 10^{32}$~ergs.  Adding in the five CMEs with kinetic energy $>$1.4$ \times 10^{32}$~ergs that occurred in 2003 (Events~\#11, 12, 13, 14, and 16 in Table~\ref{tbl-1}), we see that only 21 CMEs with kinetic energy $\gapprox 10^{32}$~erg occurred during 1997--2003. For this period, \citet{2006JApA...27..243G} reports a total of 4133~CMEs, with average kinetic energy $5\times10^{29}$~erg.  Therefore, if we consider the 1997-2003 period as typical, only $21/4133 \simeq $0.5\% of all CMEs have kinetic energies as large as those considered in this paper.  In summary, the events studied represent the largest solar eruptive events that occurred during the period in question.

Despite the relatively large number of events in our list, the range of energies is typically only about two orders of magnitude in any of the energy components. The uncertainties on the energy estimates of each component are generally at least an order of magnitude, except for the few cases where the bolometric radiated energy is measured with an accuracy of a factor of $\sim$2. Thus, in most cases, we cannot expect to see any significant trends with the size of the events over the limited range of our selected events. In spite of these selection effects and measurement limitations, the scatter plots nevertheless reveal several useful results regarding large SEEs in general and about specific events in particular. With a few notable exceptions, the points in each plot are bunched together within the expected order-of-magnitude uncertainties. However, there is substantially more scatter of the points in some parameters than others.  Only a few events stand out as outliers in certain plots; these outliers are discussed in Section~\ref{outliers}.

The general bunching of the data points in each scatter plot is characterized by the logarithmic RMS deviations parallel and perpendicular to the line of constant ratio that passes through the logarithmic centroid. The $X$ and $Y$ values of the centroids are shown on each plot and listed along with the RMS values in Table \ref{tbl-scatterparams}. The fact that the data points generally bunch together can be interpreted as an extension to SEEs of the ``\emph{big flare syndrome}'' (BFS), a phrase coined by \citet{1982JGR....87.3439K}
based on the strong correlations between proton fluxes and associated microwave and hard X-ray burst parameters, and a concept which has since come to mean that each flare component scales roughly linearly with some absolute measure of flare ``size.'' At that time, before the so-called \emph{``solar flare myth''} was exposed \citep{1993JGR....9818937G}, it was not clear that SEPs were generally more likely to be accelerated at CME shock fronts, but now the same concept can be applied to include CMEs with flares and SEPs. From our data, it is clear that, with some caveats, the BFS concept can be applied to all energetic SEE phenomena, including the CME energy.

Notable exceptions to the scatter of the points being consistent with the expected uncertainties in each parameter are the following:

\begin{itemize}

\item The plot of the GOES 1 -- 8 $\mbox{\AA}$ integrated energy vs.\ the total energy radiated from the SXR-emitting plasma over all wavelengths (Figure~\ref{goes_thermal}a) shows that these two parameters are well correlated (the Spearman's rank correlation coefficient $\rho$ given in Table~\ref{tbl-scatterparams} is 0.96, the highest for any pair of parameters). This is not surprising since the measurements of the two parameters are not independent -- they both use the same GOES X-ray data. The scatter of the points thus reflects only the range of flare sizes and the different temperatures -- the scatter perpendicular to the line of constant ratio, only about half an order of magnitude, is the result of different temperatures, while the scatter parallel to that line is almost two orders of magnitude, reflecting the range of flare intensities.  A regression analysis of the data in Table~\ref{tbl-1} shows that the ratio (best estimate $\pm$ standard error) of the total energy radiated by SXR-emitting thermal plasma in the flare to the observed GOES 1 -- 8 $\mbox{\AA}$ flux is $15.4 \pm 0.8$, i.e., that the energy radiated in the GOES 1 -- 8 $\mbox{\AA}$ band is about one-fifteenth to one-twentieth ($R=0.05$) of the total energy radiated by the SXR-emitting plasma over all wavelengths. The presence of the outlier points for Events \#3 and 8, several RMS values away from the line of constant ratio, suggests that these two events are different in that their temperature is lower than the average. This is consistent with the conclusion reached by \cite{1996ApJ...460.1034F} and \cite{2004SpWea...2.2002G} that lower temperatures are generally associated with smaller X-ray peaks.

\item The range of the ion energies -- about three orders of magnitude -- is larger than for all other parameters.  This range is especially evident in Figure~\ref{ions_electrons}a, which shows the scatter plot for the energy in flare-accelerated ions as determined from the {\em RHESSI} gamma-ray observations (Section~\ref{ions}) vs.\ the energy in flare-accelerated electrons as determined from {\em RHESSI} hard X-ray observations (Section~\ref{electrons}).  Since the ion energy content above 1~MeV was deduced by applying the same $\delta=4$ spectral extrapolation to the energy content above 30~MeV in all events, this large range in energy contents above 1~MeV mirrors exactly the spread in the energy contents above 30~MeV and hence the spread in the observed 2.223~MeV line fluences.  However, it must be noted that uncertainty in the value of $\delta$ results in a further large uncertainty in the ion energy content above 1~MeV for any specific event, since this quantity is derived through spectral extrapolation over one-and-a-half orders of magnitude in ion energy (Section 2.5). This uncertainty would act to {\it increase} the scatter in the energy content above 1~MeV if the value of $\delta$ was positively correlated with the value of the $>$30~MeV energy content; alternatively, it would act to {\it decrease} the scatter in energy content $>$1~MeV if the value of $\delta$ was negatively correlated with the value of the $>$30~MeV energy content.  In this context, it should be noted that \cite{2009ApJ...698L.152S} found a strong correlation, over more than three orders of magnitude, between the energy in flare-accelerated ions and the energy in electrons above 300~keV (rather than the $\sim$20~keV lower cutoff energy used here).  This suggests that steep [shallow] spectra (high [low] values of $\delta$) are associated with low [high] values of the energy content at high energies, and therefore that use of individual spectral indices to create more accurate energy estimates of the ion energy content above 1~MeV might reduce the scatter in the plot.

\end{itemize}

\section{Discussion of Outlier Data Points}
\label{outliers}

While most of the events in Table~\ref{tbl-1} lie (by definition) within the $2\sigma$ ellipses in the various cross-correlation plots, there are a few notable exceptions.  We now discuss these ``outliers'' and the possible reasons for their unusual energetic partitioning.

\begin{itemize}
\item Event \#1. This M5.1 event, on 2002~February~20, is one of the two events with a relatively low ratio of CME energy (kinetic + potential) to magnetic energy (Figure~\ref{cme_mag}a).  Table~\ref{tbl-1} shows a paucity of data for other energetic components.  Data for GOES 1 -- 8~$\mbox{\AA}$ emission and total radiated energy from the SXR-emitting plasma are available; the event does fall outside the $2\sigma$ ellipse in Figure~\ref{goes_thermal}a, but only by virtue of its overall weakness, not the ratio of GOES 1 -- 8 $\mbox{\AA}$ emission to total SXR-emitting energy.  We believe that this is therefore simply a weak event, in which only a small fraction of the available magnetic energy was dissipated.

\item Event \#3. This C5.0 event, on 2002~May~22, has a relatively low ($\lapprox 1\%$) ratio of GOES~1~--~8~$\mbox{\AA}$ emission to total SXR-emitting energy (Figure~\ref{goes_thermal}a).  An event with limited overall information (Table~\ref{tbl-1}), it does, however, appear in Figure~\ref{ions_electrons}b, where a normal ratio of SEP to CME energy is evident.  It should be noted that this is the only GOES C-class event in Table~\ref{tbl-1} and we therefore simply  categorize this event as a weak GOES event, possibly due to the low temperature ($<9$ MK) of the soft-X-ray-emitting plasma. This event is the third in a sequence of events starting with an M1.5 event peaking at 21:29~UT on May~21, followed by a C9.7 event at 00:30 on May~22 and the C5.0 event in question at 03:34~UT. {\em RHESSI} saw parts of each of these events and the 6-12~keV images show that they came from three distinctly different locations with the following spatial centroid coordinates (in arcseconds): (-550, 270), (880, -330), (750, -350), respectively. The peak temperatures derived from the GOES data for the three events are 13, 11, and 8~MK, respectively. Thus, this event was much cooler than the other larger events in our list.

\item Event \#8.  Data for this M4.6 event, on 2002 November 9, is available for all energetic components other than flare-accelerated ions (Table~\ref{tbl-1}); consequently, the event appears in most of the scatterplots in Figures~\ref{goes_thermal}--\ref{cme_mag}. In most of the plots, the event is situated within the general bunching of points.  However, Figure~\ref{cme_mag}b reveals a low ratio of bolometric energy to CME energy, while Figure~\ref{cme_mag}c shows a similarly high ratio of energy in accelerated particles to bolometric radiated energy.  Together, these point simply to an event with a relatively low bolometric radiance, as inferred from the FISM model (Section~\ref{Bolometric}).  This arises because of the relatively short duration of the GOES event (13 minutes from peak 1 -- 8 $\mbox{\AA}$  flux to 50\% of peak).  As pointed out by \citet{2012SoPh..279...23C}, ``the total radiated output of flares depends more on the flare duration than the typical GOES X-ray peak magnitude classification.'' The relatively low temperature of this flare (peak value of 13 MK) also explains the relatively low ratio of GOES 1 -- 8 $\mbox{\AA}$ integrated flux to thermal radiated energy (Figure~\ref{goes_thermal}a).  This event also appears in Figure~\ref{ions_electrons}b as having a marginally low ratio of SEP energy to CME kinetic energy, but this is a presumably unrelated phenomenon.

\item Event \#9. This X1.4 event, on 2003~May~27, has the lowest ratio of flare-accelerated ion energy to flare-accelerated electron energy (Figure~\ref{ions_electrons}a).  The ratios of peak thermal energy and broad-band SXR radiated energy to energy in flare-accelerated nonthermal particles are average (Figures~\ref{goes_thermal}c and~\ref{goes_thermal}d), and the ratio of GOES 1 -- 8 $\mbox{\AA}$ to total SXR-emitting energy is nominal (Figure~\ref{goes_thermal}a).  \cite{2009ApJ...698L.152S} show that this flare has a ratio of accelerated $\gtrsim$20~MeV~nucleon$^{-1}$ ions to accelerated relativistic electrons that is comparable to other gamma-ray flares, suggesting that this flare is an outlier because it has a relative deficiency in higher-energy particles, both ions and electrons.  In particular, the stated ion energy content may be a significant underestimate if the ion spectrum is steeper than the $\delta=4$ power-law assumed in the extrapolation from the $>$30~MeV proton energy value that is obtained from the observed neutron-capture line fluence.

\item Event \#12. This X17 event, on 2003~October~28, has relatively high CME and magnetic energies (Figure~\ref{cme_mag}a and Table~\ref{tbl-1}).  It was one of three events from the ``Halloween'' active region of October-November 2003, in which a very high non-potential magnetic energy value ($4 \times 10^{33}$~ergs; see Figure~\ref{ions_electrons}d) was inferred.  This event is therefore an ``outlier'' simply because it was a very large event; there are no particularly unusual ratios of energetic components.

\item Event \#17.  The data set for this X1.6 event, on 2004~July~15, is quite extensive, with all components measured except CME and SEP energies (which are understandably absent given the $\sim$E45$^\circ$ location of the flare). The event has a slightly low ratio of bolometric radiant energy to magnetic energy (Figure~\ref{ions_electrons}d), and slightly high ratios of to flare-accelerated particle energy to bolometric radiated energy (Figure~\ref{cme_mag}c) and energy radiated by SXR-emitting plasma to bolometric radiated energy (Figure~\ref{cme_mag}d).  It has the equal-lowest bolometrically radiated energy content, and {\it the} lowest thermal energy content of SXR-emitting plasma, of any event studied (Table~\ref{tbl-1}), but interestingly does {\it not} show as an outlier in any other plots in which it appears, notably the plot of thermal energy content vs. thermal radiated energy (Figure~\ref{goes_thermal}b) and the plot of thermal energy content vs. accelerated particle energy (Figure~\ref{goes_thermal}c) (although it is barely inside the 2$\sigma$ ellipse in both of these plots).  As with Event \#8, these event properties may be explained by the relative short SXR duration of this flare - only 4 minutes from the GOES 1 -- 8 $\mbox{\AA}$ peak to 50\% of peak flux. Although the peak plasma temperature derived from the GOES fluxes was 22 MK, the temperature stayed above 15 MK for only about 8 minutes. The short duration and low temperatures lead the FISM modeling process (Section~\ref{Bolometric}) to assign a correspondingly low estimate of the bolometric radiant energy \citep{2012SoPh..279...23C}.

\item Event \#18.  Data for this M7.1 event, on 2004~July~25, is rather limited (Table~\ref{tbl-1}).  The event appears as an outlier in Figure~\ref{ions_electrons}d, due to a low ratio of bolometric energy to magnetic energy. Table~\ref{tbl-1} shows that this event had one of the highest inferred non-potential magnetic energy contents of all the events studied, while Figure~\ref{goes_thermal}a shows very small values of both radiation in the GOES 1 -- 8 $\mbox{\AA}$ waveband and total energy radiated by the SXR-emitting plasma (although the ratio of the two is nominal).  {\em RHESSI} did show a flare flag starting at 05:37~UT, but entered night at 05:42~UT, three minutes after the GOES start time; therefore no reliable electron or ion energy measurements are available.  Further, no CME was observed for this event, which was located at approximately W30$^\circ$ longitude.  As a result, it is not known whether the event was simply very weak radiatively, or whether only a relatively small part of the nonpotential magnetic energy available was released.

\item Event \#25. For this X7.1 event, on 2005~January~20, the ratios of nonthermal particle energy to bolometric radiated energy (Figure~\ref{cme_mag}c), soft-X-ray-radiated to peak thermal energy (Figure~\ref{goes_thermal}b), and soft-X-ray-radiated to nonthermal particles (Figure~\ref{goes_thermal}d) are nominal.  However, the ion to electron energy ratio ($\sim$5:1; Figure~\ref{ions_electrons}a and Table~\ref{tbl-1}) is conspicuously high; indeed, this event had one of the highest ion energy contents measured.  This high ion-to-electron energy ratio may in part be due to the presence of high-energy protons (inferred from the 2.223~MeV line) for several minutes after most of the electron-ion bremsstrahlung had dissipated.  The event also has a very low ratio of CME energy to magnetic energy (Figure~\ref{cme_mag}a), but a very high ratio of SEP energy to CME energy (Figure~\ref{ions_electrons}b). The ratio of bolometric radiated energy to free magnetic energy is nominal (Figure~\ref{ions_electrons}d), but it is the only event with a ratio of bolometric radiated energy to CME energy that is greater than 100\% (Figure~\ref{cme_mag}b). Together, these results show that the main component that makes this event an ``outlier'' is the low CME energy. Note, however, that the CME kinetic energy in this event is very uncertain.

The reasons for these unusual circumstances are not completely understood.  Much has been written about this event \citep[e.g.,][and references therein]{2008SoPh..252..149G} but the acceleration of the intense flux of SEPs has still not been fully resolved. It has been suggested that in this event the SEPs were accelerated at the flare site rather than in the CME shock \citep{2005AIPC..781..246L,2005AdSpR..35.1857L}. In similar vein, \cite{2006A&A...445..715S, 2007A&A...472..309S} concluded that, ``the relativistic protons were not accelerated by the CME-driven shock.'' The event produced a cosmic-ray ground-level enhancement that is among the largest recorded in the history of cosmic ray measurements \citep{2011AdSpR..48.1232M}. However, based on the particle spectrum measured over a wide rigidity range (1 - 20~GV), \cite{2008ICRC....1..285M} state that, ``The 2005 January 20 GLE was an unusual event in its intensity and brevity, placing it on the outer edges of parameter space for shock acceleration to GeV energies, but still not requiring a different process, i.e., direct solar-flare acceleration.'' Similarly, \cite{2009ApJ...693..812R} finds that SEPs in this event, as in other ground-level events (GLEs), are produced after the shock onset and don't require a separate non-shock injection. On the other hand, \cite{2011SSRv..tmp...99M} conclude that neutron monitor observations of this event indicate two separate pulses of high-energy particles, one accelerated by the flare, and a second accelerated by the CME-driven shock.

\item Event \#30. This M6.4 event, on 2005~August~25, had a very low peak thermal energy content and a relatively low ratio of the total SXR-emitted energy to the peak thermal energy content (Figure~\ref{goes_thermal}b). It also has a low ratio of total SXR-emitted energy to nonthermal particle energy (Figure~\ref{goes_thermal}d).  No CME or SEP data are available for this East-limb event.  For this event, the GOES flux decayed from the M6.4 peak to the C6 level (10\% of the peak flux) in only $\sim$14 minutes, so that the low values of the radiated energy in SXR-emitting plasma are due to this simply being a relatively short-lived event.

\end{itemize}

\section{Conclusions}
\label{conclusions}

\begin{figure}
\begin{center}
   {\includegraphics[width=0.9\textwidth]{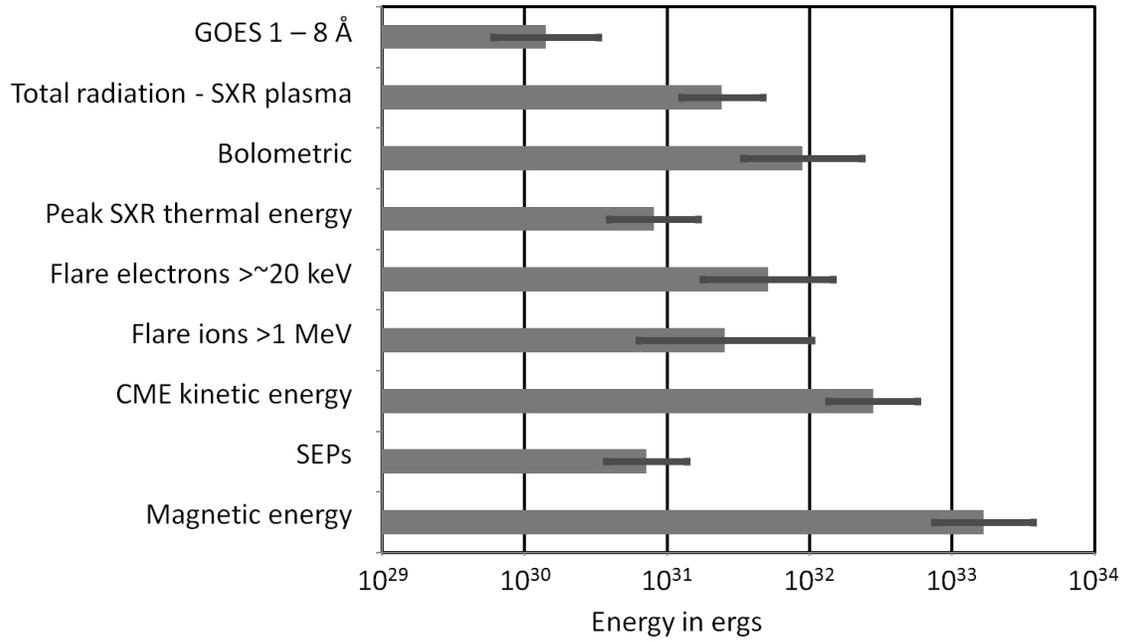}}
\end{center}
    \caption[]{\small{Bar chart showing the (logarithmic) average energies of the different components for the six events for which values were obtained for all components -- Events \#13, 14, 20, 23, 25, and 38. The short thin bars show the $\pm1\sigma$ logarithmic scatter of the energies of the six events.}
}
    \label{histogram}
\end{figure}

Despite the rather large uncertainties in the individual measurements used in this analysis, the relatively large number (38) of events nevertheless allows us to reach some general conclusions about the ``typical'' ratios of various energetic components in large SEEs.  We have found the following general statements to hold:

\begin{itemize}
  \item Figure \ref{goes_thermal}b shows that the total energy radiated by the SXR-emitting plasma over the course of the event exceeds, by about half an order of magnitude ($R = 2.8$ in Table \ref{tbl-scatterparams}), the peak energy content of the thermal plasma that produces this radiation.  This reinforces the conclusions of \cite{1980sfsl.work..341M} that some form of energy is continuously supplied to this hot plasma throughout the event;

  \item Figures \ref{goes_thermal}d and \ref{cme_mag}c show that the energy content in flare-accelerated particles (electrons and ions) is sufficient to create not only the total energy radiated by the SXR-emitting plasma, but also the total bolometric radiated energy of the event;

  \item Figure \ref{ions_electrons}a shows that the energy contents of flare-accelerated ions and electrons are comparable at the order-of-magnitude level. This result supports the earlier claims of \cite{1995ApJ...455L.193R} and \cite{2000IAUS..195..123R} and has significant consequences for acceleration models;

  \item Figure \ref{ions_electrons}b shows that the SEP energy is typically a few percent ($R = 0.04$ in Table \ref{tbl-scatterparams}) of the CME kinetic energy in the solar wind rest frame, a result with implications for shock-acceleration models of interplanetary particles;

  \item The combination of Figures \ref{ions_electrons}d, \ref{cme_mag}a, and \ref{cme_mag}c shows that the available magnetic energy is indeed sufficient to power the thermal plasma, flare-accelerated particles, and the CME. Although some ``double-counting'' may be involved in summing these energy components \citep[e.g., both the flare-accelerated particles and the CME may transfer energy to the ambient plasma; see][]{2005JGRA..11011103E}, this result nevertheless conforms to the widely-held view that the source of the energy released in SEEs lies in stressed magnetic fields.
\end{itemize}

Figure \ref{histogram} shows the logarithmic average (i.e., the geometric mean) of the energies with $\pm1\sigma$ logarithmic scatter of the various energy components for the six events (events \#13, 14, 20, 23, 25, 38) for which {\it all} energetic components were measured. (Events \#6 and \#12 were not included, since for these events some of the components were determined only as upper or lower limits -- see Table~\ref{tbl-1}.) This figure, coupled with the overall ratios summarized in Table~\ref{tbl-scatterparams}, succinctly demonstrates how, in very approximate terms, the available magnetic energy gets distributed in a ``typical'' flare in our sample:

\begin{enumerate}

\item Of the $\sim$$10^{33}$ ergs of available non-potential magnetic energy, approximately 30\% is released in the SEE, with the remainder staying in the active region as stored magnetic energy.  Of the $\sim$30\% that is released, some 80\% ($\sim$25\% of the available energy) is released in the CME (mostly as kinetic energy) and approximately 20\% ($\sim$5\% of the available energy) is released as flare-accelerated particles, roughly evenly distributed between electrons and ions.

\item All of the energy in the flare-accelerated particles appears to ultimately emerge as radiation across a wide range of wavelengths, from optical to soft X-rays \citep{2005JGRA..11011103E}.
However, only about one-third of the energy in flare-accelerated particles ($\sim$2\% of the available stored energy) is ultimately radiated from high-temperature soft X-ray-emitting plasma.  The maximum amount of energy stored as enhanced thermal energy in the soft X-ray-emitting plasma is $\sim$1\% of that released, and the amount of energy radiated in the diagnostic GOES 1--8 \AA\ waveband is only about 5\% of the total energy radiated by the SXR-emitting plasma, or $\sim$0.1\% of the available magnetic energy.

\item Because of the need for a CME to ``overtake'' the solar wind and form a shock front where SEPs can be accelerated, only about two-thirds of the kinetic energy carried by the CME ($\sim$15\% of the available nonpotential magnetic energy) is available for SEP acceleration.  The SEP production process is in turn $\sim$4\% efficient, so that only about half a percent of the released magnetic energy ultimately appears in the form of SEPs.

\end{enumerate}

Although for completeness we have listed Spearman's rank correlation coefficient ($\rho$) values in Table \ref{tbl-scatterparams}, little significance can be attached to these values other than for the obviously tight correlations between parameters that are essentially derived from the same data (e.g., the bolometric emission, the energy radiated from the SXR-emitting plasma, and the energy radiated in the GOES 1 -- 8 \AA\ band, all of which are dependent on GOES SXR flux measurements).  Any correlations amongst independent components are masked by the large uncertainties in the individual measurements used in the various scatter plots. Progress in this direction will require sampling of events over a much larger range in flare size to determine if the distribution of energies amongst the different components found here for large events is preserved for smaller events, as would be expected for a ``\emph{big SEE syndrome}.'' Such a project is the next step towards a more comprehensive understanding of energy release in SEEs.

Using data from the new, more sophisticated, instruments that are now available will allow more accurate energy estimates to be made of some of the components. For example, the detailed differential emission measure analysis now possible using data from the EUV Imaging Spectrometer \citep[EIS;][]{2007SoPh..243...19C} on Hinode, and from the Atmospheric Imaging Assembly \citep[AIA;][]{2012SoPh..275...17L} and the Extreme Ultraviolet Variability Experiment \citep[EVE;][]{2012SoPh..275..115W} on the Solar Dynamics Observatory (SDO), should lead to better estimates of the energies in the thermal plasma. Vector magnetograms from the Helioseismic and Magnetic Imager \citep[HMI;][]{2012SoPh..275..207S} on SDO allow for more accurate estimates of the energy in the nonpotential magnetic field.

Other energy components not considered in our analysis may be found to contain significant total energy and should be included in any future compilation of global energetics. These include the turbulent mass motions revealed by the broadening of atomic lines seen with EUV and X-ray spectrometers \citep[e.g.,][]{2008uxss.book.....P}, and the cumulative heating of CME plasma reported by \cite{2011ApJ...735...17M} to be comparable to (or even greater than) the CME kinetic energy. Another aspect not discussed here is the question of a second flare phase that, according to \cite{2011ApJ...739...59W} and \cite{2011ApJ...731..106S}, can release a similar amount of energy as in the initial phase.  Nevertheless, we believe that the order-of-magnitude comparisons of energetic components presented herein represents a significant advance in our understanding of the nature of energy release in SEEs.

\acknowledgments

We thank Gordon Holman, Richard Schwartz, and Kim Tolbert for help with analyzing the GOES and {\em RHESSI} data, and Anil Gopie for doing most of the GOES data analysis.  We also thank the referee for an unusually comprehensive and thorough review of the originally submitted version of this manuscript, which resulted in a significantly improved paper.  AGE was supported by NASA Grant NNX10AT78J, RAM by NASA grants NNX08AI11G and NNX11AO75G, and AV by various NASA grants to the Naval Research Laboratory.  SoHO is a joint ESA and NASA mission. CHIANTI is a collaborative project involving researchers at NRL (USA), RAL (UK), and the Universities of Cambridge (UK), George Mason (USA), and Florence (Italy).

\appendix
\section{Appendix}
\label{appendix}

Here, for definiteness, we provide the equations used to determine the values listed in Table~\ref{tbl-scatterparams}: the (logarithmic) centroid energies $X_{centroid}$ and $Y_{centroid}$, their ratio $R$, the RMS ($1\sigma$) values both perpendicular and parallel to the lines of constant ratio, and the Spearman's rank correlation coefficient $\rho$.

The coordinates of the logarithmic centroid in each plot are given by

\begin{eqnarray}
\log_{10} X_{centroid} & = & \frac{1}{N}\sum_{i=1}^N \, \log_{10}X_i  \,\,\,; \nonumber \\
\log_{10} Y_{centroid} & = & \frac{1}{N}\sum_{i=1}^N \, \log_{10}Y_i \,\,\, ,
\label{eq_centroid}
\end{eqnarray}
where $N$ is the number of events for which there are viable measures of both components included in the scatter plot in question.

The lines of constant ratios ($R = 100$\%, 10\%, and 1\%) between the $X$ and $Y$ components satisfy the following relation:

\begin{eqnarray}
    R & = & Y/X \nonumber \\
    or \log_{10}R & = &  \log_{10}(Y/X) \nonumber \\
                  & = & \log_{10} Y - \log_{10} X \,\,\, .
    \label{eq_R}
\end{eqnarray}
Lines of constant logarithmic average event energy ($\log_{10}A$), shown in the plots as dashes every order of magnitude along the lines of constant ratio, are defined as follows:

\begin{eqnarray}
    \log_{10}A & = & (\log_{10}X + \log_{10}Y)/2 \nonumber \\
               & = & \log_{10} \sqrt{XY} \,\,\, .
    \label{eq_A}
\end{eqnarray}
The RMS deviations of the points perpendicular to and parallel to the line of constant ratio passing through the centroid are defined by

\begin{eqnarray}
    RMS_{\perp} & = & (1/N) \sqrt{\sum_i \, (\log_{10} R_i - \log_{10}R_{centroid})^2} \nonumber \\
                & = & (1/N)\sqrt{\sum_i \, (\log_{10} (Y_i/X_i) - \log_{10}(Y_{centroid}/X_{centroid}))^2}
     \label{eq_RMSperp}
\end{eqnarray}
and

\begin{eqnarray}
    RMS_{\parallel} & = & (1/N)\sqrt{\sum_i \, (\log_{10} A_i - \log_{10}A_{centroid})^2} \nonumber \\
                    & = & (1/2N)\sqrt{\sum_i \, [\log_{10} (X_i Y_i) - \log_{10} (X_{centroid} Y_{centroid})]^2} \,\,\, .
    \label{eq_RMSparal}
\end{eqnarray}
These RMS values were used to draw the ellipse with axes of $2\times$RMS$_{\perp}$ and $2\times$RMS$_{\parallel}$ around the logarithmic centroids in each plot.

The Spearman's rank correlation coefficient $\rho$ is calculated by first assigning ranks $x_i \, (=1,\ldots,N)$ and $y_i \, (=1,\ldots,N)$ to the $X$ and $Y$ values, respectively, of the $N$ points used in the plot in question. The ranks are assigned such that $X(x_i) \le X(x_{i+1})$, $Y(y_i) \le Y(y_{i+1})$, with ``ties'' assigned the average rank of the tied values.  (Note that the rank order does not depend on whether we use the $X_i$ values or their logarithms $\log_{10} X_i$.)  The Spearman's rank correlation coefficient is then calculated as the correlation coefficient of the ranks:

\begin{equation}\label{spearman}
\rho={\sum_{i=1}^N (x_i - {\overline x}) (y_i - {\overline y}) \over
\sqrt{\sum_{i=1}^N (x_i - {\overline x})^2 \sum_{i=1}^N (y_i - {\overline y})^2}} \,\,\, ,
\end{equation}
where ${\overline x}$ and ${\overline y}$ are the means of the ranks $x_i$ and $y_i$, respectively.  For a monotonic dependence of $Y$ on $X$, $\rho=1$, even if the variables do not obey a perfect linear correlation.

\bibliographystyle{apj}	
\bibliography{emslie_et_al}

\end{document}